\documentclass[letterpaper,twocolumn,10pt]{article}
\usepackage{usenix2019_v3}

\usepackage[subtle]{savetrees}

\usepackage{comment}
\usepackage[makeroom]{cancel}

\usepackage{lipsum}
\usepackage{titlesec}

\titlespacing\section{0pt}{11pt plus 4pt minus 2pt}{7pt plus 2pt minus 2pt}
\titlespacing\subsection{0pt}{10pt plus 4pt minus 2pt}{3pt plus 2pt minus 2pt}
\titlespacing\subsubsection{0pt}{9pt plus 4pt minus 2pt}{3pt plus 2pt minus 2pt}

\usepackage{graphicx}
\usepackage{subfig}
\usepackage{color}
\usepackage[shortlabels]{enumitem}
\usepackage{authblk}

\usepackage{verbatim}
\usepackage{amsmath}
\usepackage{amssymb}
\usepackage{pifont}

\newcommand{\xmark}{\text{\ding{55}}}

\DeclareMathOperator*{\argmin}{arg\,min}

\newcommand{\bestmodelacc}{99.74\%}
\newcommand{\bestmodelfpr}{0.56\%}

\newcommand*{\red}[1]{\textcolor{red}{#1}}

\newcommand{\authnote}[2]{{\bf \textcolor{blue}{#1}: \em \textcolor{red}{#2}}}
\newcommand{\yizheng}[1]{\authnote{Yizheng}{#1}}

\begin{document}

\date{}

\renewcommand\Affilfont{\itshape}
\title{\Large \bf On Training Robust PDF Malware Classifiers}

{
\author{Yizheng Chen}
\author{Shiqi Wang}
\author{Dongdong She}
\author{Suman Jana}
\affil{\vspace{-8pt}Columbia University}
}



\maketitle
\begin{abstract}
Although state-of-the-art PDF malware classifiers can be trained with
almost perfect test accuracy (99\%) and extremely low false positive rate (under 0.1\%),
it has been shown that even a simple adversary can evade them.
A practically useful malware classifier must be robust against evasion attacks. However, achieving
such robustness is an extremely challenging task. 

In this paper, we take the first steps towards training robust PDF malware classifiers with verifiable robustness 
properties. For instance, a robustness property can enforce that no matter how many pages from
benign documents are inserted into a PDF malware, the classifier must still classify it as malicious. We demonstrate 
how the worst-case behavior of a malware classifier with respect to specific robustness properties can be formally verified. Furthermore, we find that training classifiers that satisfy formally verified robustness properties 
can increase the evasion cost of unbounded (i.e., not bounded by the robustness properties) attackers by eliminating simple evasion attacks.


Specifically, we propose a new distance metric that operates on the PDF tree structure
and specify two classes of robustness properties including subtree insertions and deletions. We utilize
state-of-the-art verifiably robust training method to build robust PDF malware classifiers.
Our results show that, we can achieve 92.27\% average verified robust accuracy over three properties, while maintaining 
\bestmodelacc{} accuracy and \bestmodelfpr{} false positive rate.
With simple robustness properties, our robust model maintains 7\% higher robust accuracy than
all the baseline models against unrestricted whitebox attacks.
Moreover, the state-of-the-art and new adaptive evolutionary attackers need up to
10 times larger $L_0$ feature distance and 21 times more PDF basic mutations
(e.g., inserting and deleting objects) to evade
our robust model than the baselines.
\end{abstract}

\section{Introduction}

Machine learning classifiers have long been used for many important security problems such as malware detection, spam filtering, and online fraud detection. One of the most ubiquitous applications is to detect PDF malware, which is a very popular infection vector for both large-scale mass and targeted attacks. Many prior research projects have demonstrated that machine-learning-based PDF malware classifiers can achieve almost perfect test accuracy (99\%) with extremely low false positive rate (under 0.1\%)~\cite{smutz2012malicious, vsrndic2013detection}.
Nonetheless, all the state-of-the-art classifiers, including the proprietary ones used by popular services like Gmail,
can be evaded by trivial transformations over the PDFs, such as adding
invisible document metadata, deleting the length indicator of the exploit payload, or simply
increasing the length of the document~\cite{laskov2014practical,xu2016automatically,evade_gmail}.

Since any security-relevant application of machine learning classifiers must deal with adaptive adversaries, it is fundamentally insufficient to evaluate security classifiers by measuring the accuracy and false positive rate.
Despite the abundance of available metrics given by well-established theoretical results
in machine learning~\cite{sklearn_metrics}, none of them are suitable to measure the robustness of the classifiers under adaptive attackers. In order to be practically useful, a malware classifier must be demonstrated to be secure against different types of adaptive attacks. For example, a sample robustness property might require that no matter
how many pages from benign documents are inserted into a PDF malware, the classifier still must classify the modified malware as malicious. Similarly, deletion of any non-functional objects in the PDF must not result in a benign classification.

Ideally, a classifier should be \emph{sound} with regard to a robustness property.
That is, the robustness property can be formally verified to get strict bounds on the worst-case
behavior of the classifier. If a classifier satisfies the robustness property,
the strongest possible attacker bounded by the specification of the property,
\emph{i.e., bounded attacker},
will not be able to violate the property, no matter how powerful the attacker is
or whatever adaptive strategy she follows. For example, even for a perfect knowledge
attacker, any creative way of inserting pages from the most-benign documents to the malware
can be verified to keep the malicious classification.

If we train classifiers to be verifiably robust against building block attacks, we can raise
the bar for more sophisticated attacks to succeed. Essentially, the attacker is solving a search problem to find an evasive PDF malware. She starts from a malicious PDF,
performs a series of manipulations to the PDF, and eventually arrives at a solution that
makes the PDF variant classified as benign without affecting its maliciousness. To maintain malicious functionality,
the PDF variant needs to have the correct syntax and correct semantics. Therefore, 
manipulations from different attacks can be decomposed to many building block operations
in the parsed PDF tree. By training building block robustness properties, we eliminate simple and easy evasions, which increases the search cost for attackers.


In this paper, we take the first steps towards training a PDF malware classifier with verifiable robustness properties,
and we demonstrate that such classifiers also increase the attack cost even for the attackers not bounded by these properties. We address several challenges in building robust PDF malware classifiers. First, previous work has shown that
retraining the malware classifier with adversarial instances drastically increases the false positive rate~\cite{advtrain_pdf_slides, grosse2016adversarial}.
Since verifiably robust training is strictly a harder problem to
solve than adversarially robust training without any verifiable bound, it is challenging to specify robustness properties that do not
increase false positive rates yet still increase the cost for the attackers. To this end,
we propose a new distance metric for the structured PDF trees.
Using a small distance for the robustness properties maintains low false positive rate.
Second, popular model choices for PDF malware classifiers are not suitable
for robust training. For example, adversarially
robust training over a random forest model requires manual adjustment to the complexity of
trees to maintain acceptable accuracy~\cite{kantchelian2016evasion}.
Therefore, we choose a neural network model to leverage
state-of-the-art verifiably robust training schemes.
Third, to evaluate our defense, we compare the robustness of our models
against twelve different baseline models using seven attacks.
We implement five attacks unrestricted by robustness properties, including
feature-space attacks as well as application-space attacks that generate actual evasive PDF malware.
In particular, we develop adaptive attacks to target the trained robustness properties based on
EvadeML~\cite{xu2016automatically}.
We use these attacks to quantify the increase in the unbounded attacker cost caused by the verifiable robust training.


Using our new distance metric for the PDF tree structure, we specify two classes of robustness properties,
subtree deletion properties and subtree insertion properties. The properties
allow any possible attacks involving deletion/insertion up to a bounded number of subtrees under the PDF root.
For example, when choosing to delete
\texttt{/Root/Metadata} subtree containing children \texttt{/Root/Metadata/Length} and 
\texttt{/Root/Metadata/Type}, the attacker can delete either one of the children, both children,
or the whole subtree.
Note that even at the subtree distance one, the properties include a large number of
possible model inputs. For example,
subtree insertion property bounds the attacker to any one of the 42 subtrees under the PDF root.
Among them, \texttt{/Root/Pages} alone includes $2^{1,195}$ different input features
for the classifier. This overapproximates attacker's actions, and includes even unknown attacks.
We train seven verifiably robust models with different robustness properties,
utilizing symbolic interval analysis~\cite{reluval2018, shiqi2018efficient}.
We measure the verified robust accuracy (VRA) for
a test set of 3,416 PDF malware. The VRA represents the percentage of test samples that
are verifiably robust against the strongest bounded attacker.
Although adversarially robust training is known to achieve strong robustness against
a specific type of attacker, the gradient attacker~\cite{madry2017towards},
our verifiably trained models can obtain superior verifiable robustness against \emph{all} possible bounded attackers
while keeping high test accuracies and low false positive rates.


Perhaps even more importantly, we show that a verifiably robust classifier with two proposed robustness properties can already increase the cost for the \emph{unbounded} attacker.
We evaluate our model against two unrestricted whitebox attacks and three unrestricted blackbox attacks.
In the whitebox setting, our robust model maintains 7\% higher estimated robust accuracy
(defined in Section~\ref{sec:ERA and VRA}) against the unrestricted gradient attack and
the Mixed Integer Linear Program (MILP) attack than the baseline models.
In the blackbox setting, the enhanced evolutionary attack needs up to 3.6 times larger $L_0$ distance
and 21 times more PDF mutations
(described in Section~\ref{section:evademl_implementation} and~\ref{sec:Adaptive Evolutionary Attacker})
to evade our model compared to the baselines.
Even the adaptive evolutionary attack needs 10 times larger $L_0$ distance and 3.7 times more PDF mutations to evade our robust model.
In addition, we achieve 2\% higher ERA than the strongest baseline model against the reverse mimicry attack.
The results show that training verifiably robust PDF malware classifiers even for carefully chosen simple robustness properties can effectively increase the bar for the attacker to solve the evasion problem.




As defenders, making all evasion attacks on malware classifiers computationally infeasible is an extremely hard problem.
However, our work shows a very promising direction to increase the cost of an attacker by training malware classfiers that are verifiably robust against different simple robustness properties.
We can potentially further increase the robustness of PDF malware classifier by specifying
more complex robustness properties.
Our key contributions are summarized as follows.

\begin{itemize}[leftmargin=*]
\itemsep0em
\item We are the first to evaluate and train verifiable robustness properties of PDF malware classifiers.
We propose a new distance metric to bound the robustness properties in the PDF tree structure.
We specify two essential robustness properties as building blocks to compose more powerful properties. 
\item We train verifiably robust PDF malware classifier models.
We thoroughly evaluate the robustness against twelve baseline models,
using state-of-the-art measures including
estimated robust accuracy (ERA) under gradient attacks and verified robust accuracy (VRA) against any bounded adaptive attacker. We can achieve 92.27\% average VRA over three robustness properties while maintaining \bestmodelacc{} test accuracy and \bestmodelfpr{} false positive rate.
\item We can increase the bar
for unrestricted attackers to evade our verifiably robust model.
Our model achieves 7\% higher ERA
against the unrestricted gradient attacker up to 200,000 iterations than all the baselines.
The state-of-the-art and new adaptive evolutionary attackers need up to
10 times larger $L_0$ feature distance and 21 times more PDF manipulations to evade
our robust model.



\vspace{-5pt}

\end{itemize}

\section{Background}
\label{sec:Background}

\begin{figure*}[t!]
	\centering
        \subfloat[Example objects in a PDF malware.]
        {\includegraphics[scale=0.35]{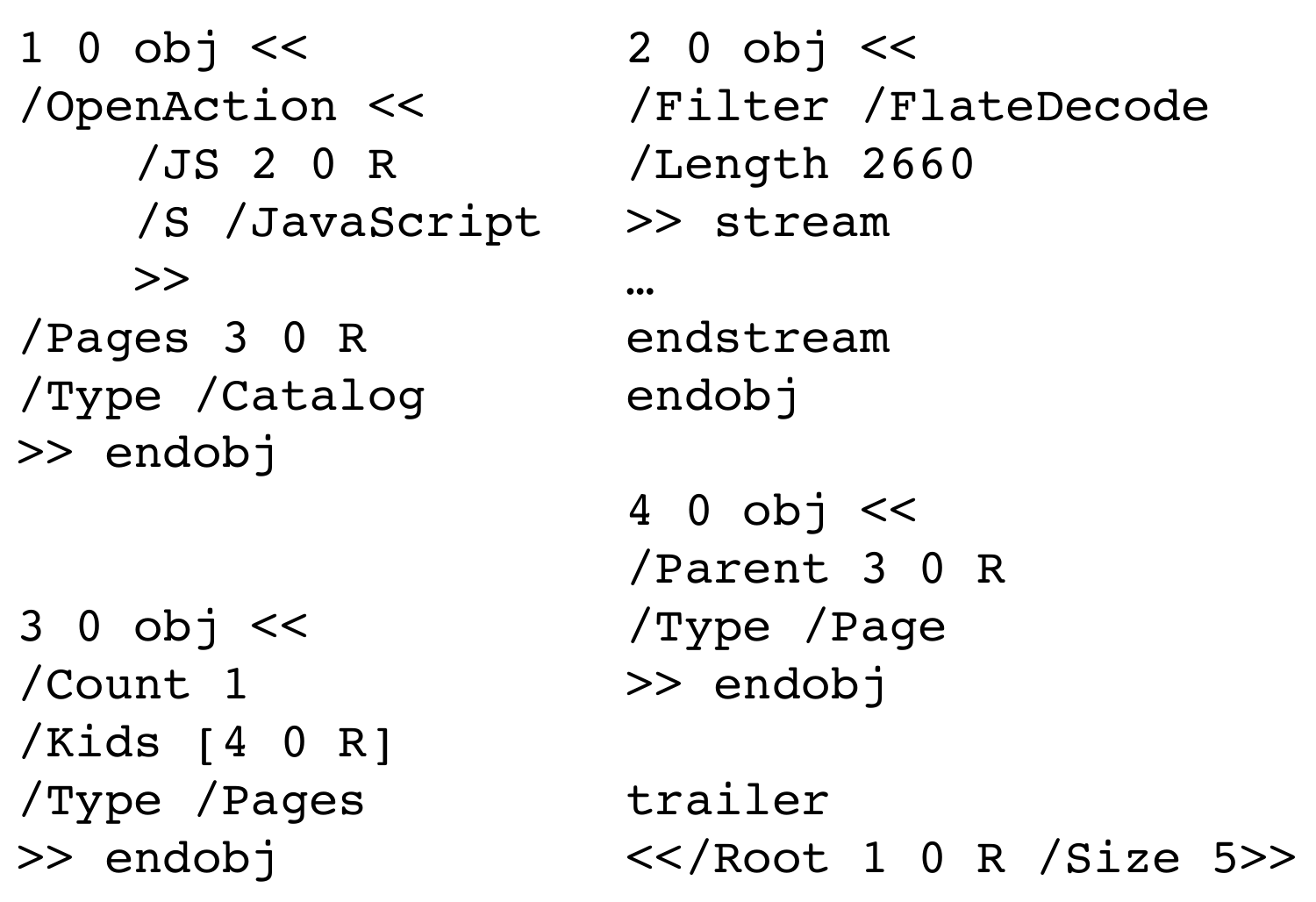}
        \label{fig:pdf_objects}}
	~
        \subfloat[The tree structure of a PDF malware.]
        {\includegraphics[scale=0.35]{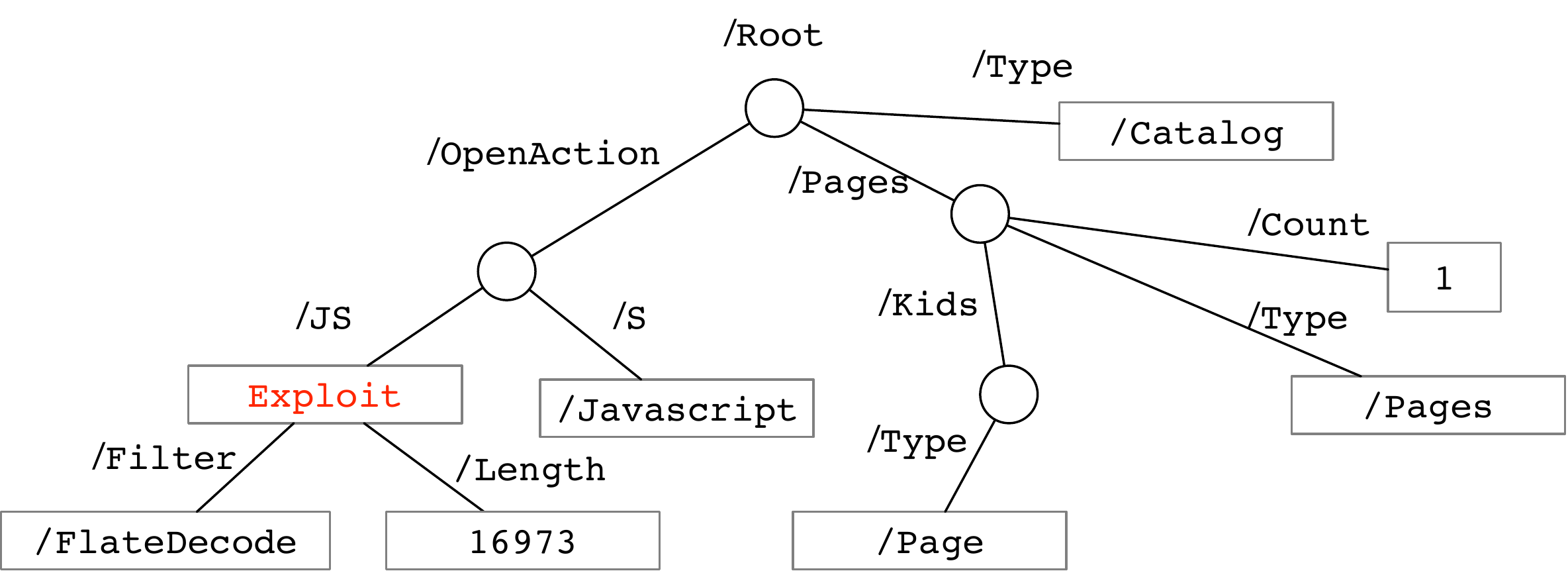}
        \label{fig:pdf_tree}}
	~
        \subfloat[Hidost features.]
        {\includegraphics[scale=0.35]{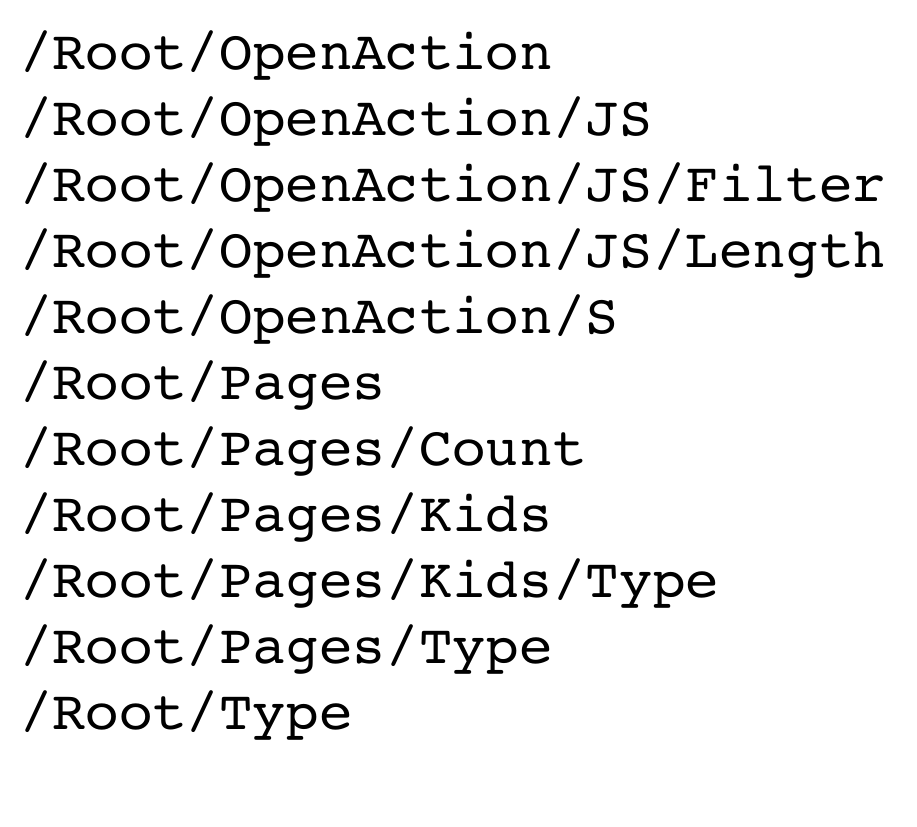}
        \label{fig:pdf_feat}}
        \caption{ The objects and trailer, parsed PDF tree structure, and extracted Hidost features
        from an example PDF malware.
        }
        \label{fig:pdf_example}
\end{figure*}

In this section, we present an overview of the PDF format and PDF malware. Then, we introduce the features used by PDF malware classifiers and two main classes of attacks that evade them. At last, we describe the robust training techniques.

\subsection{PDF Malware}
\label{sec:PDF Malware}

The Portable Document Format (PDF) contains four sections: header, body, cross-reference table, and trailer.
The header identifies the file format, version,
and a magic number.
The body consists of various types of objects, including arrays,
name trees, dictionaries, etc.
For example, Figure~\ref{fig:pdf_objects} shows four PDF objects and the trailer from a PDF malware.
The trailer identifies the entrance to parse the file, along with the cross-reference table size. Here, the entrance is the root object \texttt{1 0 obj}, where the object number is 1
and the object version is 0, and \texttt{R} means indirect reference.
The cross-reference table indexes all object locations in the file.
Starting from the root object,
a parser resolves referred objects either using the cross-reference table or scanning the PDF to get the object locations. 

The four objects in this file are dictionaries, indicated by $<<$ and $>>$ symbols and
enclosed by \texttt{obj} and \texttt{endobj} keywords. The dictionary object is a collection of
key/value pairs. Each key is a name object, and the value can be any object.
The root object \texttt{1 0 obj} has a special type, \texttt{/Catalog}, and the value of the key \texttt{/OpenAction}
is another dictionary object. Within \texttt{/OpenAction}, the object containing the JavaScript exploit
is referred to as \texttt{2 0 R}. The exploit object contains a stream that can be decoded using the \texttt{/Filter} indicator, and a length field for the stream. The exploit is executed when the file is opened. 
There is generally discrepancy between
the parser implementation and actual file format specification.
For example, many PDF readers do not need the correct
length field to decode the stream, and malware authors can delete the field to evade the classifier.
The rest of the PDF contains object 3 and 4 that refer to each other.
The PDF structure forms a tree, by taking the
shortest path to objects via references (Figure~\ref{fig:pdf_tree}).


PDF malware exploits the vulnerabilities in the PDF reader in order to transfer execution control,
e.g., to run shellcode or drop additional binary.
PDF malware authors employ various techniques to evade the detection, e.g., add content from legitimate
documents, crash the PDF reader, and obfuscate the PDF content.
Making PDF malware classifier robust against trivial manipulation remains a hard problem. For example, increasing the length of the file to be 7,050,000 bytes can evade the Gmail PDF malware scanner~\cite{evade_gmail}.


%




\subsection{PDF Malware Classifiers}
\label{sec:PDF Malware classifiers}
In this section, we discuss two open-source PDF malware classifiers that have attracted
considerable evasion effort in the security community,
PDFrate~\cite{smutz2012malicious} and Hidost~\cite{vsrndic2013detection}.


\subsubsection{PDFrate}

PDFrate~\cite{smutz2012malicious} uses a total of 202 features including counts for various keywords and certain fields in the PDF. For example, number of characters in the author field,
number of ``endobj'' keyword, sum of all pixels in all the images, and number of JavaScript markers, etc.
The classifier is a Random Forest, with 99\% accuracy and 0.2\% false positive rate over the
Contagio malware dataset~\cite{contagio2011}.

Simple manipulation of the PDF file can result in very big changes in the feature values of PDFrate. For instance, inserting pages from a benign document to
the PDF malware can increase the page count feature alone to be as big as the maximal integer value,
which also affects many other counts.
If a bounded manipulation in the PDF cannot tightly bound
the feature input to the classifier, these features are not suitable for verifiably robust training.

\subsubsection{Hidost}

Hidost~\cite{vsrndic2013detection} uses Bag-of-Path features extracted from the parsed tree structure of the PDF.
It obtains the shortest structural path to each object, including terminals and non-terminals in the tree,
and uses binary counts for these paths as features. In the paper, the authors used only
those paths that appeared in at least 1,000
files in the corpus, which reduced the number of paths from 9 million to 6,087.
Hidost was evaluated on a decision tree model and a SVM model. Both models have
99.8\% accuracy and less than 0.06\% false positive rate

The binary Bag-of-Path features are able to bound the input to the classifier, given certain attack properties.
For example, in our dataset, if we insert anything under the \texttt{/Pages} subtree,
only up to 1,195 features will be flipped from 0 to 1, resulting in a tight input bound to the classifier.
Therefore, in this paper, we choose to use Hidost features to build our robust PDF malware classifier.

\subsubsection{Automatically Evading Malware Classifiers}
\label{section:Automatically Evading Malware Classifiers}

Several automated attacks have successfully evaded PDF malware classifiers, under different threat models.

\textbf{White-box Attacks.} White-box attackers are assumed to have perfect knowledge.
They can launch
precise attacks targeting the exact model being trained, e.g., gradient-based attack~\cite{biggio2013evasion, laskov2014practical}.
For instance, in the white-box setting, the Gradient Descent and Kernel Density Estimation (GD-KDE) attack
can be launched against the SVM version of PDFrate~\cite{laskov2014practical}.
In addition, ~\cite{grosse2016adversarial} uses an approach to only add features, in order to preserve existing malicious functionality of adversarial malware examples~\cite{arp2014drebin}.
The drawback of such white-box gradient-based attacks is that the evasion instances are found
in the feature space, so they do not generate actual PDF malware.





\textbf{Black-box Attacks.}
The threat models of black-box attacks generally assume that the attacker does not have access
to any model parameters, but has oracle access to the prediction labels for some samples,
and in some cases also the prediction confidence.
In some settings, features and the model type are assumed to be known as well.
Xu et al.~\cite{xu2016automatically} use a genetic evolution algorithm to automatically evade
both PDFrate and Hidost. The evolutionary algorithm uses a fitness score
as feedback, to guide the search in finding evasive PDF variants by mutating the seed PDF malware.
For every generation of the population during the search, the attack uses a cuckoo oracle to dynamically check that
mutated PDFs still preserve the malicious functionality. This check is much stronger
than static insertion-only methods used by gradient-based attacks.
Dang et al.~\cite{dang2017evading} uses a more restricted threat model where the attacker does
not have access to classification scores, and only has access to the classified label and a blackbox morpher
that manipulates the PDFs. They use the scoring function based on Hill-Climbing to attack the classifier
under such assumptions. In this paper, we improve the attack
from the genetic evolution framework of EvadeML~\cite{xu2016automatically},
and also develop several adaptive attacks based on that.




\subsection{Robust Training}

Out of the arms race between adversarial image examples~\cite{szegedy2013intriguing}
and many defenses~\cite{papernot2015distillation, papernot2017practical,carlini2017towards,athalye2018obfuscated,cao2017mitigating}, two training methods
have proven to be the strongest among all. They are adversarially robust training and verifiably robust
training. We briefly explain the training methods, and reason about why verifiably
robust training provides stronger robustness guarantee.

\subsubsection{Robust Optimization}
\label{sec:Robust Optimization}

Both adversarially robust training and verifiably robust training are based on robust optimization.
Let us first look at the optimization objective used by the regular training process of the neural network.
Given an input $x$ with the true label $y$, a neural network $f_{\theta}$ parameterized by $\theta$
maps it to a label $\hat{y} = f(x)$. A loss function $L(y, \hat{y})$ is used to evaluate the errors
of such prediction, e.g., the cross-entropy loss. The training process has the following optimization
objective that minimizes the loss to find optimal weights $\theta$. The summation is an empirical measure
of the expected loss over the entire training dataset.

\begin{equation} \label{eq:1}
\theta = \argmin_{\theta} \sum{L(y, \hat{y})}
\end{equation}

In the adversarial setting, for the input $x$, there can be
a set of all possible manipulations $\tilde{x}$ bounded by a distance metric $D_k$ within distance k,
i.e. $\tilde{x} \in D_k{(x)}$. Robust optimization minimizes the worst case loss for all inputs in $D_k{(x)}$,
solving a minimax problem with two components.

\begin{equation} \label{eq:2}
\theta = \argmin_{\theta} \sum{ \max_{\tilde{x} \in D_k{(x)}} L(y, f_{\theta}(\tilde{x}) ) }
\end{equation}

\begin{itemize}
\item {\bf Inner Maximization Problem:} find $\tilde{x}$ that maximizes the loss value within the robustness
region $D_k{(x)}$, i.e., the \emph{robust loss}.
\item {\bf Outer Minimization Problem:} minimize the maximal loss to update the weights $\theta$ of the neural network.
\end{itemize}

The following two robust training methods solve the inner maximization problem in different ways.

\subsubsection{Adversarially Robust Training}

Adversarially robust training empirically estimates the maximal loss in Equation~\ref{eq:2}
by using different attacks. The state-of-the-art adversarially robust training method from Madry et al.~\cite{madry2017towards} 
uses adversarial examples found by the Projected Gradient Descent (PGD) attack~\cite{kurakin2016adversarial}
to estimate the robust loss for the training.  The training method has been applied to benchmarking image datasets, including MNIST and CIFAR-10. The trained models have shown robustness against 
known attacks including the Projected Gradient Descent (PGD) attack~\cite{kurakin2016adversarial},
Carlini-Wagner (CW) attacks~\cite{carlini2017towards}, Fast Gradient Sign Method (FGSM)~\cite{goodfellow2014explaining},
etc.

Adversarially robust training has been applied to malware datasets.
In the followup work to~\cite{xu2016automatically}, Xu et al.~\cite{advtrain_pdf_slides}
applied adversarially robust training over the Contagio malware dataset, 
which increased the false positive rate to as high as $85\%$.
Grosse et al.~\cite{grosse2016adversarial} applied the training method to
the android malware classifier using adversarial malware examples, increasing
the false positive rate to $67\%$.

\subsubsection{Verifiably Robust Training}

Verifiably robust training uses sound over-approximation techniques to obtain the upper bound of
the inner maximization problem. Different methods have been used to formally verify the robustness
of neural networks over input regions~\cite{lomuscio2017approach,katz2017reluplex,huang2017safety,ehlers2017formal,fischetti2017deep,dutta2018output,raghunathan2018certified}, such as abstract transformations~\cite{gehrai}, symbolic interval analysis~\cite{reluval2018, shiqi2018efficient}, convex polytope approximation~\cite{wong2018provable}, semidefinite programming~\cite{raghunathan2018semidefinite}, mixed integer programming~\cite{tjeng2017evaluating}, Lagrangian relaxation~\cite{dvijotham2018dual} and relaxation with Lipschitz constant~\cite{weng2018towards}, which essentially solve the inner maximization problem. By using the worst case bounds derived by formal verification
techniques, verifiably robust training~\cite{dvijotham2018training,mirman2018differentiable,wong2018scaling,wang2018mixtrain} can obtain such verified robustness properties.

The training method has been applied to image datasets to increase verifiable robustness,
usually with the tradeoff of lower accuracy and
higher computation and memory cost for the training. Recent works have focused on
scaling the training method to larger networks and bigger datasets~\cite{wong2018scaling, wang2018mixtrain}.
Since verifiably robust training techniques can train classifiers to be \emph{sound} with regard to the
robustness properties, the trained network
gains robustness against even unknown adaptive attacks.
On the contrary, adversarially robust training is limited by the specific threat model
used to generate adversarial instances for the training. Therefore,
we apply verifiably robust training to build the PDF malware classifier. By carefully specifying
useful robustness properties, our robust model has only \bestmodelfpr{} false positive rate.

\subsubsection{ERA and VRA}
\label{sec:ERA and VRA}

In this paper, we will use the following two metrics to evaluate our verifiably robust PDF malware classifier.

\textbf{Estimated Robust Accuracy (ERA)} measures the percentage of test inputs that are robust against known attacks, given a distance bound.
For instance on the MNIST dataset, Madry et al.'s training method~\cite{madry2017towards}
can achieve 89\% ERA against PGD attacks within a bounded distance of  $L_{\infty} \leq 0.3$.

\textbf{Verified Robust Accuracy (VRA)} measures the percentage of test inputs that are
verified to be correctly classified within a distance bound. For example, Wong et al.'s training method~\cite{wong2018scaling}
obtains 21.78\% VRA on a CIFAR10 Resnet model within a bounded distance of $L_{\infty} \leq 8/255$.

\section{Verifiably Robust PDF Malware Classifier}

Since it is extremely hard, if not impossible, to have a
malware classifier that is robust against all possible attackers, we aim to train classifiers to be robust
against building block attacks. In this section, we describe the specification and training of robustness properties.

\subsection{Robustness Properties}
\label{sec:Robustness Properties}

\subsubsection{Motivation}

\textbf{Building block operations.}
A large group of evasion attacks against malware classifiers can be considered as solving a search problem,
e.g., mimicry attacks~\cite{wagner2002mimicry, laskov2014practical}, EvadeML~\cite{xu2016automatically}, EvadeHC~\cite{dang2017evading} and MalGAN~\cite{hu2017generating}.
The search starts from the seed malware, modifies the malware to generate variants
while keeping the malicious functionality, until finding a variant that can be classified as benign.
The attacks use building block operations to make the search process
more systematic and efficient over a large space. Specifically for PDF malware, the operations include
PDF object mutation operators~\cite{xu2016automatically}, random morpher~\cite{dang2017evading}
and feature insertion-only generator~\cite{hu2017generating}.
After performing the building block operations, the attacks optimize the search
based on the classifier's feedback that indicates the evasion progress.
We want to make the search harder by training classifiers to be robust against
building block operations. To achieve that, we consider operations
that generate PDFs with the correct syntax. A PDF variant needs to have both correct syntax
and correct semantics to stay malicious.
Though dynamic execution can confirm the same malicious behavior, it is too expensive to do that during training. Therefore, we statically ensure the correct PDF syntax.
A syntactically correct
PDF file can be parsed into a tree structure (Figure~\ref{fig:pdf_tree}, Section~\ref{sec:PDF Malware}).
Thus, the building block operations are a combination of insertion
and deletion in the PDF tree. Based on this insight, we design robustness properties related
to the PDF subtrees. We propose two classes of subtree insertion and subtree deletion properties,
which can be used as the stepping stones to construct more sophisticated attacks.




\textbf{False positive rate.}
It is crucial to maintain low false positive rate for security classifiers due to the
Base-Rate Fallacy~\cite{axelsson1999base}. If we train classifiers with evasive
malware variants without enforcing a proper bound, the classifier will have
very high false positive rate. Since attacks often mimic benign behavior,
the feature vectors of unbounded variants are close to benign vectors, which affects the false positive rate.
Therefore, we need a distance metric to define the robustness properties
to capture the similarity between the PDF malware and its variants.
Since the $L_p$ norm distance in the feature space does not capture whether the corresponding
PDF variant has the correct syntax, we propose a new distance metric
for the PDF subtree.


\subsubsection{Subtree Distance Metric}



\textbf{Subtree Distance Definition.}
We propose a new distance metric to bound the attacker's building block operations over
a PDF malware.
The subtree distance between two PDFs $x$ and $x'$ is, the number
of different subtrees of depth one in the two PDF trees. These subtrees
are directly under the root object in the PDF,
regardless of their height and the number of nodes in them. Formally,

$d(x, x') = \#\{(\text{root}_x.\text{subtrees} \cup \text{root}_{x'}.\text{subtrees}) - (\text{root}_x.\text{subtrees} \cap \text{root}_{x'}.\text{subtrees}) \}$

We first take the union of the subtrees with depth one from two PDFs, and then remove the intersection of the two subtree sets (identical subtrees). The distance $d(x, x')$ is the cardinality of the resulting set.


If the attacker inserts benign pages into the PDF malware
under the \texttt{/Root/Pages} subtree (Figure~\ref{fig:pdf_tree}), this operation will not exceed subtree distance one, no matter how long the malicious PDF document becomes.
Changing an arbitrary subtree in the PDF may have different $L_p$ norm distances depending
on which subtree is manipulated.
For example, in the Hidost binary path features, manipulating \texttt{/Root/Metadata} is bounded by $L_1 \leq 4$,
whereas changing \texttt{/Root/Pages} can be up to $L_1\leq 1195$. However, under
the subtree distance, they are both within the distance one bound.

We use the subtree distance to define robustness properties. Each property corresponds to an
over-approximated set $D_k(x) = \{\tilde{x} | d(x, \tilde{x}) \leq k \}$. The set captures all PDF malware $\tilde{x}$ that can be possibly generated by changes in arbitrary $k$ subtree regions under the root of the malware seed $x$, regardless of the feature extraction method.
Since insertion and deletion
are building block operations, we formulate
these robustness properties at distance one before composing more complicated
robustness properties.

\subsubsection{Subtree Insertion and Deletion Properties}

\textbf{Subtree Insertion Property (Subtree Distance 1):} given a PDF malware, all possible manipulations to the PDF
bounded by inserting an arbitrary subtree under the root, do not result in a benign prediction by the classifier.

The attacker first chooses any one of the subtrees,
and then chooses an arbitrary shape of the subtree
for the insertion. Some subtrees are commonly seen in benign PDFs, which can be good insertion candidates
for evasion, e.g., \texttt{/Root/Metadata}, \texttt{/Root/StructTreeRoot},
\texttt{/Root/ViewerPreferences}. Although the subtree distance for the property is only one,
the total number of allowable insertions is on
the order of the sum of exponentials for the number of children under each subtree.

The property over-approximates the set of semantically correct
and malicious PDFs. For example, if we insert
\texttt{/Root/Names/JavaScript/Names/JS} but not \texttt{/Root/Names/JavaScript/Names/S},
the javascript is no longer functional. Moreover, we over-approximate the attacker's possible actions.
Attacks are usually based on some optimization procedure rather than exhaustive search. However,
if a known attack fails to find succesful insertion in a subtree, unknown attacks may succeed.
Therefore, the property can overestimate the worst case behavior of the classifier.





\textbf{Subtree Deletion Property (Subtree Distance 1):} given a PDF malware, all possible manipulations to the PDF
bounded by deleting an arbitrary subtree under the root, do not result in a benign prediction by the classifier.


For the PDF malware example shown in Figure~\ref{fig:pdf_tree}, this property allows deleting any one of the following: \texttt{/Root/Type}, \texttt{/Root/Pages}, and
\texttt{/Root/OpenAction}. Note that this allows any combination of deletion under non-terminal nodes \texttt{/Root/Pages} and \texttt{/Root/OpenAction}.

Some exploit triggers may be lost or the program semantics may be broken by deleting content from the malware. 
The robustness property covers an over-approximated set of evasive PDF malware, and enforces that
they are always classified as malicious. It is acceptable to include some non-malicious PDFs
in the robustness region, as long as we do not increase the false positive rate for benign PDFs.



%

\subsubsection{Other Properties}

We do not specify other common properties like replacement, since many can be viewed as a combination of insertions and deletions. The robustness properties can be generalized to up to $N$ subtree distance, where $N=42$ in our feature space.
Next, we describe properties with larger distances.



\textbf{Subtree Deletion Property (Subtree Distance 2):}
the strongest possible attackers bounded by deletions within any two subtrees under the root,
cannot make the PDF classified as benign.

\textbf{Subtree Insertion Property (Subtree Distance $N-1$):}
the strongest possible attackers bounded by insertions within all but one subtree under the root,
cannot make the PDF classified as benign.


\textbf{Monotonic Property and Subtree Insertion Property (Distance $N$):}
Incer et al.~\cite{incer2018adversarially} have proposed to enforce the monotonic property for malware classifiers.
The monotonic property states that an attacker cannot evade the classifier by only increasing the feature values.
Specifically, if two feature vectors satisfy $x \leq x'$, then the classifier $f$ guarantees that $f(x) \leq f(x')$.
They enforce monotonicity for both benign and malicious classes, such that inserting features into any executable makes it appear more malicious to the classifier.
The property is so strong that it decreased the temporal detection rate of the classifier by 13\%.

To compare against the monotonic property, we propose the subtree insertion property at distance $N$.
In other words, the insertion is unrestricted by any subtree, and it is allowed for all features.
We focus on this property for the malicious PDFs, which is a key difference from the monotonic property.

Larger distances bound a larger set of evasive malware variants, which can make
malicious feature vectors more similar to benign ones and affect the false positive rate.
In our evaluation, we train all five properties and several combinations of them
using mixed training technique (Table~\ref{tab:vra}).

\subsection{Training the Properties}

Given the over-approximated set of inputs $D_k(x)$ for each robustness property, we use sound analysis of
the neural network to obtain the corresponding robust loss.

\textbf{Sound analysis definition.}
A sound analysis over the neural network $f_\theta$ represents a sound transformation $T_f$ from the input to the output of $f_\theta$. Formally, given input $x \in \mathbb{X}$ and a property $D_k(x)$ bounded by distance $k$, the transformation $T_f$ is sound if the following condition is true: $\forall x \in \mathbb{X}$, we have 
$\{f_\theta(\tilde{x})|\tilde{x} \in D_k{(x)}\}\subseteq T_f(D_k{(x)})$
That is, the sound analysis over-approximates all the possible neural network outputs for the property.
Using $T_f(D_k{(x)})$, we can compute the robust loss in Equation~\ref{eq:2}.

\textbf{Training.}
Existing works have shown that training only for the robustness objective
degrades regular test accuracy, and combining the two objectives helps smooth the
conflict between the two~\cite{mirman2018differentiable, wang2018mixtrain, gowal2018effectiveness}.
Consequently, we adopt the same principle to train for a combined loss as below.

\begin{equation} \label{eq:3}
L = L(y, f_\theta(x)) + \max_{\tilde{x} \in D_k{(x)}} L(y, f_{\theta}(\tilde{x}))
\end{equation}

In Equation~\ref{eq:3}, the left-hand side of the summation denotes the regular loss for the training data
point $(x, y)$, and the right-hand side represents the robust loss for any manipulated $\tilde{x}$ bounded
by distance $k$ satisfying a defined robustness property $D_k{(x)}$. We give the same weights to combine the two parts of the loss, to equally optimize the regular loss and the robust loss.
The robust loss is computed by the worst case within the bounded region of every training data input. More implementation details about robust training can be found in Section~\ref{sec:Verifiably Robust Models}.



\section{Evaluation}
\label{sec:Evaluation}

\begin{table*}[ht!]
  \centering
  \footnotesize
  \begin{tabular}{llrrr | rrr rrrr | r}
    \hline
     & & & & & & \multicolumn{5}{c}{Knowledge and Access} & &  \multicolumn{1}{r}{Bounded} \\
     & \multicolumn{2}{r}{\textbf{Realizable}} & & & \multicolumn{3}{c}{\textbf{Model}} & \textbf{Training} & \multicolumn{2}{c}{\textbf{Classification}} & \textbf{Knows} & {by} \\
    & \textbf{Attacker} & \textbf{Input} & \textbf{Type} & \textbf{Eval} & \textbf{Feat} & \textbf{Arch} & \textbf{Wgts} & \textbf{Alg, Data} & \textbf{Label} & \textbf{Score} & \textbf{Defense}  & \textbf{Property}\\
    \hline
     (1) & Bounded Arbitrary Attacker &  \checkmark & I & VRA & \checkmark & \checkmark & \checkmark & \checkmark & \checkmark & \checkmark & \checkmark & \checkmark \\
     (2) & Bounded Gradient Attacker &  \xmark & I & ERA & \checkmark & \checkmark & \checkmark & \checkmark & \checkmark & \checkmark & \checkmark & \checkmark \\
     \hline
     \hline
     (3) & Unbounded Gradient Attacker & \xmark & II & ERA & \checkmark & \checkmark & \checkmark & \checkmark & \checkmark & \checkmark & \checkmark & \xmark \\
     (4) & MILP Attacker & \xmark & II & ERA & \checkmark & \checkmark & \checkmark & \checkmark & \checkmark & \checkmark & \checkmark & \xmark \\
     (5) &Enhanced Evolutionary Attacker & \checkmark & II & ERA & \checkmark & \checkmark & \xmark & \xmark & \checkmark & \checkmark & \xmark & \xmark  \\
     (6) & Reverse Mimicry Attacker & \checkmark & II & ERA & \checkmark & \checkmark & \xmark & \xmark & \checkmark & \checkmark & \xmark & \xmark \\
     (7) & Adaptive Evolutionary Attacker & \checkmark & III & ERA & \checkmark & \checkmark & \xmark & \xmark & \checkmark & \checkmark & \checkmark & \xmark \\
    \hline
  \end{tabular}
  \caption{We evaluate our models using seven types of attackers.
  They represent, two strongest bounded adaptive attackers (Type I),
  four state-of-the-art unbounded attackers (Type II),
  and the new adaptive unbounded attacker (Type III).
  Only attackers (1) and (2) are restricted by the robustness property. The gradient and MILP attackers ((2), (3), (4)) operates on non-realizable feature-space inputs. The other attackers operate on realizable inputs, among which attacker (1) overapproximates realizable inputs.}
  \label{tab:threat_model}
\end{table*}

We train seven verifiably robust models and compare them against twelve baseline models,
including neural network with regular training, adversarially robust training,
ensemble classifiers, and monotonic classifiers\footnote{Our code is available at https://github.com/surrealyz/pdfclassifier}. We answer the following questions
in the evaluation.
\begin{itemize}
\itemsep0em
\item Do we have higher VRA and ERA if the attackers are restricted by the robustness properties?
\item Do we have higher ERA against unrestricted attackers?
\item How much do we raise the bar (e.g., $L_0$ distance in features and mutation trace length) for the unrestricted attackers to evade our robust models?
\end{itemize}

\begin{figure}[t]
    \centering
    \scalebox{0.38}{\includegraphics{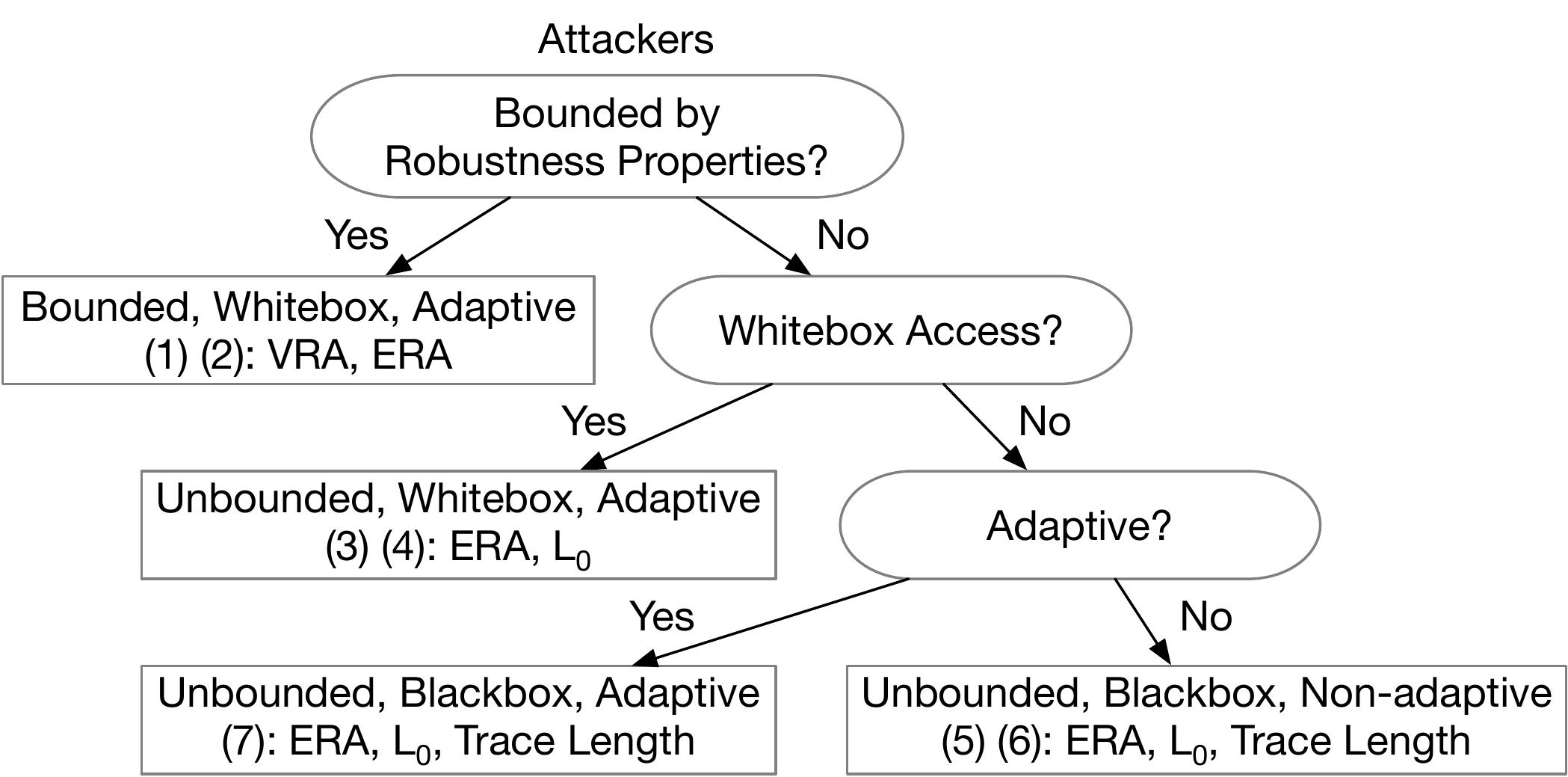}}
    \caption{
    Different types of attackers in our evaluation.
    }
   \label{fig:hierarchy}
\end{figure}

We use seven different attackers to evaluate the models.
When choosing the attackers, we consider three factors, 
i.e., whether the attacker is bounded by the robustness properties, whether the attacker
has whitebox access to the model, and whether the attacker is adaptive.
Figure~\ref{fig:hierarchy} shows the categories where every attacker belongs to,
and the evaluation metrics we use for the category. The detailed threat model
for each attacker ((1) to (7)) is shown in Table~\ref{tab:threat_model}.
In the table, we have marked whether each attacker generates realizable inputs that are real PDF malware. We evaluate attacks
producing both realizable and non-realizable inputs since robustness
against them are equally important. Tong et al.~\cite{tong2019improving} have shown that robustness against feature-space attacks on non-realizable inputs can be generalized to robustness against realizable attacks.



\textbf{Machine.} We use a desktop machine for all the experiments. The machine is configured with
Intel Core i7-9700K 3.6 GHz 8-Core Processor, 64 GB physical memory, 1TB SSD,
Nvidia GTX 1080 Ti GPU, and it runs a 64-bit Ubuntu 18.04 system.
To run the genetic evolution attacks, we set up the Cuckoo sandbox with 32 virtual machines running
Windows XP SP3 32 bit and Adobe Acrobat Reader 8.1.1.

\subsection{Models}


\subsubsection{Datasets}

\begin{table}[t!]
  \begin{center}
  	\small
    \begin{tabular}{lcc}
      \hline
      \textbf{Dataset} & \textbf{Training PDFs} & \textbf{Testing PDFs} \\
      \hline
      Malicious &  6,896 & 3,448 \\
      Benign    & 6,296 & 2,698  \\
      \hline
      \end{tabular}
    \caption{The Contagio~\cite{contagio2013} dataset used for regular training.}
    \label{tab:dataset}
  \end{center}
\end{table}

We obtain the PDF malware dataset from Contagio~\cite{contagio2013}.
The malicious PDFs include web exploit PDFs and email attachments for targeted attacks.
We split the dataset into 70\% train set and 30\% test set, summarized in Table~\ref{tab:dataset}.
In total, we have 13K training PDFs and 6K test PDFs.
We use the Hidost feature extractor to extract structural paths features, with
the default \texttt{compact} path option~\cite{vsrndic2013detection, hidost_feat}.
The input features have 3,514 dimensions, representing all the distinct
path features from the training dataset.

\textbf{Robustness Properties.}
In our experiments, we focus on five robustness properties as labeled from A to E in Table~\ref{tab:intervals}.
For brevity, we will refer to the four robustness properties as property A (subtree deletion distance one), B (subtree insertion distance one),
C (subtree deletion distance two), D (subtree insertion distance 41) and E (subtree insertion distance 42).
They are defined in Section~\ref{sec:Robustness Properties}.
Every subtree is represented by a continuous range of indices in the binary feature vector, so
insertion and deletion can be bounded by a corresponding interval.

\begin{table}[t!]
  \begin{center}
  \small
    \begin{tabular}{lrr}
     \hline
     & \textbf{Training} & \textbf{Testing} \\
      \textbf{Robustness Property} & \textbf{Intervals} & \textbf{Intervals} \\
      \hline
      \textbf{A:} Subtree Deletion Distance 1 &  30,655 & 15,672 \\
      \textbf{B:} Subtree Insertion Distance 1    & 288,414 & 143,472  \\
      \textbf{C:} Subtree Deletion Distance 2   & 62,445  &  33,711 \\
      \textbf{D:} Subtree Insertion Distance 41    & 288,414 & 143,472 \\
      \textbf{E:} Subtree Insertion Distance 42  & 6,867 & 3,416 \\
      \hline
      & \textbf{Training} & \textbf{Testing} \\
      \hline
      Number of PDF Malware    & 6,867 & 3,416 \\             
      \hline
      \end{tabular}
    \caption{Five robustness properties and the corresponding number of intervals used to train
    and test VRA.
    The intervals are extracted from 6,867 training and 3,416 testing PDF malware.}
    \label{tab:intervals}
  \end{center}
\end{table}

\textbf{Symbolic Interval Analysis.}
We implement verifiably robust training using Symbolic Interval Analysis~\cite{reluval2018, shiqi2018efficient} as the sound over-approximation method.
Symbolic interval analysis uses intervals to bound the adversarial input range to the model,
then propagates the range over the neural network while keeping input dependency.
When passing the interval through the non-linearity in the neural network, we do linear relaxation of the input interval, and bound the output interval by two equations~\cite{shiqi2018efficient}. The analysis tightly over-estimates the output range, which we use to compute the robust loss (Equation~\ref{eq:2}, Section~\ref{sec:Robust Optimization}).

We parse and manipulate the PDFs using the modified version of pdfrw parser that handles malformed PDF malware~\cite{weilin_pdfrw}.
When a subtree is deleted, the paths it contains and objects with zero reference are deleted.
If the subtree contains any object
referenced by other nodes, the object still exists along with the other paths.
Within the regular training dataset, we have successfully parsed 6,867 training and 3,416 testing PDF malware to train and evaluate the robustness properties.
Table~\ref{tab:intervals} shows the number of intervals we extract for each property, separated by training and testing sets.

\subsubsection{Model Architecture and Hyperparameters}
\label{sec:Model Architecture and Hyperparameters}

Among the models we evaluate, many are neural networks or have neural networks as the basic component.
We use the same model architecture and training hyperparameters for all the neural network models.
We follow previous work~\cite{grosse2016adversarial, guo2018lemna, saxe2015deep} to build
a feed-forward neural network with two hidden layers, each with 200 neurons activated by ReLU, and
a final layer of Softmax activation.
We train all neural network models for 20 epochs, using the Adam Optimizer,
with mini-batch size 50, and learning rate 0.01.

\subsubsection{Baseline Models}
\label{sec:Baseline Models}

\textbf{Baseline Neural Network.}
We train the baseline neural network model with the regular training objective (Equation~\ref{eq:1}, Section~\ref{sec:Robust Optimization}), using the regular training dataset (Table~\ref{tab:dataset}).
The model has 99.9\% test accuracy and 0.07\% false positive rate. The performance is consistent
with those reported in PDFrate~\cite{smutz2012malicious} and Hidost~\cite{vsrndic2013detection}
(Section~\ref{sec:PDF Malware classifiers}).

\textbf{Adversarially Robust Training.}
We use the new subtree distance metric
to adversarially retrain the neural network. We train five models corresponding to
A, B, C, D, and A+B properties. For the deletion properties, we train with deleting one or two entire subtrees;
and for the insertion properties, we train with inserting one or 41 full subtrees. The resulting performance
of the models are shown in Table~\ref{tab:vra}. All models have more than 99\% accuracy.
The Adv Retrain A, B models maintain the same 0.07\% FPR as the baseline neural network.
The other three models have slightly higher FPR up to 0.15\%.

\textbf{Ensemble Classifiers.}
Ensemble methods use multiple learners to boost the performance of the base learner.
We implement two ensemble classifiers.

\textbf{Ensemble A+B.}
The provable robustness property
is, if a PDF variant is generated by subtree insertion bounded by distance one to
a PDF, the variant has the same prediction as the original PDF.
The ensemble classifies the PDF as malicious, if an arbitrary full
subtree deletion results in malicious class by the base learner.
We augment the regular training dataset with an arbitrary subtree deleted from
both malicious and benign PDFs, which maintains the performance
for original PDFs because they also need to be tested under multiple deletion operations.
Ensemble A+B achieves 99.87\% accuracy and 0.26\% FPR.

\textbf{Ensemble D.} The provable robustness property is,
if a PDF variant is generated by inserting up to 41 subtrees
in a PDF, it has the same prediction as the original PDF.
For the base learner, we train
a neural network to classify the original malicious and benign PDFs as if they were generated
by up to 41 subtree insertions. Consequently, we augment the training dataset by keeping
one subtree from all PDFs to train the base learner. To build the ensemble, we test every single
subtree of the unseen PDFs, and predict the malicious class if any subtree is classified as malicious.
The Ensemble D model
has 99.96\% accuracy and 0.07\% FPR.

\textbf{Monotonic Classifiers.}
Monotonic classifiers~\cite{incer2018adversarially} are the most related work to ours.
We follow Incer et al.~\cite{incer2018adversarially}
to use Gradient Boosted Decision Trees~\cite{chen2016xgboost} with monotonic constraint.
After comparing different tree depths, we find that the results do not significantly differ in this dataset.
Therefore, we train multiple monotonic classifiers with different number of learners (10, 100, 1K, and 2K),
where each learner is a decision tree of depth 2. The classifiers are named by the number of
learners they have (Table~\ref{tab:vra}). The monotonic classifiers have an average of 99\% accuracy and
under 2\% FPR, which shows much better performance in a small Contagio dataset compared to results in~\cite{incer2018adversarially}. Since monotonic property
is such a strong enforcement for the classifier's decision boundaries, the malware classifier in~\cite{incer2018adversarially} has
62\% temporal detection rate over a large scale dataset containing over 1.1 million binaries.

Note that the ensembles and monotonic classifiers are the only models we train
with properties for both malicious and benign PDFs. For all the other models,
we train properties for only malicious PDFs.

\subsubsection{Verifiably Robust Models}
\label{sec:Verifiably Robust Models}


\paragraph{Robust Training.}
We train seven verifiably robust models and name them with the properties
they are trained with.
We use the same model architecture and the same set of hyperparameters
as the baseline neural network model (Section~\ref{sec:Model Architecture and Hyperparameters}).
During training, we optimize the sum of
the regular loss and the robust loss in each epoch, as defined in Equation~\ref{eq:3}.
At the mini-batch level, we randomly mix batches belonging to different properties.
For example, to train the Robust A+B model, we do mixed training for regular accuracy, the insertion property, and the deletion property alternately by mini-batches, in order to obtain two properties as well as high accuracy in the same model.

The left side of Table~\ref{tab:vra} contains the test accuracy (Acc), false positive rate (FPR),
and training time for the models.

\textbf{Training Time.}
The robust models with insertion properties
(Robust B, Robust D, Robust A+B, Robust A+B+E)
took more than an hour to train, since they have significantly
more intervals (Table~\ref{tab:intervals}) than deletion properties. On the contrary, Robust A and Robust C
models can be trained by 11 and 25 minutes, respectively. The average training time for each mini-batch is 0.036s.
Efficiently scaling the number of training points, input dimension, and network size can be achieved by
techniques from~\cite{gowal2018effectiveness, wang2018mixtrain, wong2018scaling}.

\textbf{Accuracy and FPR.}
All the robust models, except the Robust D model, can maintain
over 99\% test accuracy while obtaining verifiable robustness. Robust D model dropped the test accuracy
only a little to 98.96\%. Training robustness properties increased the false positive rates by under 0.5\% for
Robust A, B, and A+B models, which are acceptable.
For models C and D, the false positive rates increased to 1.04\% and 2.3\% respectively.
Models with property E (Robust E and Robust A+B+E), have FPR 1.93\% and 1.89\%, similar to those of the monotonic
classifiers.  
The false positive rate increases more for the insertion properties (B and E) than
the subtree deletion property (A).
The FPR is also larger for a bigger distance under the same type of operation
(C vs A, and D vs B). 

\begin{table*}[ht!]
  \centering
  \small
  \begin{tabular}{llrrr | rrrrr}
    \hline
    & & & & & \multicolumn{5}{c}{\textbf{Gained Verified Robust Accuracy (VRA, \%)}} \\
    & & & & & \multicolumn{1}{c}{Property A} & \multicolumn{1}{c}{Property B} & \multicolumn{1}{c}{Property C} & \multicolumn{1}{c}{Property D} & Property E \\
   & \textbf{Acc} & \textbf{FPR} & \textbf{Train} & \textbf{Trained} & \multicolumn{1}{c}{Distance: 1} & \multicolumn{1}{c}{Distance: 1} & \multicolumn{1}{c}{Distance: 2} & \multicolumn{1}{c}{Distance: 41} & \multicolumn{1}{c}{Distance: 42} \\
    \textbf{Model} & \textbf{(\%)} & \textbf{(\%)} & \textbf{(m)} & \textbf{Prop.} & \textbf{Subtree Del.} & \textbf{Subtree Ins.} & \textbf{Subtree Del.} & \textbf{Subtree Ins.}  & \textbf{Subtree Ins.}\\
    \hline
    \hline
    Baseline NN & 99.95 & 0.07 & <1 & None & 90.25 & 0 & 49.82 & 0 & 0 \\
    \hline
    \hline
    Adv Retrain A & 99.95 & 0.07 & 1 & A & 99.24 & 0 & 84.20 & 0 & 0 \\
    Adv Retrain B & 99.95 & 0.07 & 8 & B & 85.50 & 0 & 38.20 & 0 & 0 \\
    Adv Retrain C & 99.93 & 0.11 & 2 & C & 99.21 & 0 & 88.91 & 0 & 0 \\
    Adv Retrain D & 99.93 & 0.11 & 14 & D & 93.47 & 0 & 50.00 & 0 & 0 \\
    Adv Retrain A+B & 99.92 & 0.15 & 7 & A,B & 98.51 & 0 & 87.47 & 0 & 0 \\
    \hline
    \hline
    Ensemble A+B & 99.87 & 0.26 & 2 & A,B & 97.22 & 99.97$^*$ & 7.43 & 0 & 0 \\
    Ensemble D & 99.95 & 0.07 & 2 & D & 0 & 0 & 0 & 0 & 0 \\
    \hline
    \hline
    Monotonic 10 & 98.91 & 1.89 & <1 & E & 5.74$^*$ & 98.91$^*$ & 0$^*$ & 98.91$^*$ & 98.91$^*$ \\
    Monotonic 100 & 99.04 & 1.78 & 1 & E & 7.67$^*$ & 99.04$^*$ & 0$^*$ & 99.04$^*$ & 99.04$^*$ \\
    Monotonic 1K & 99.06 & 1.78 & 4 & E & 8.78$^*$ & 99.06$^*$ & 0$^*$ & 99.06$^*$ & 99.06$^*$ \\
    Monotonic 2K & 99.06 & 1.78 & 8 & E & 8.78$^*$ & 99.06$^*$ & 0$^*$ & 99.06$^*$ & 99.06$^*$ \\
    \hline
    \hline
    Robust A & 99.84 & 0.33 & 11 & A & 99.77 & 0 & 89.43 & 0 & 0 \\
    Robust B & 99.72 & 0.59 & 102 & B & 46.43 & 99.77 & 1.26 & 0 & 0 \\
    Robust C & 99.54 & 1.04 & 25 & C & 99.94 & 0 & 99.77 & 0 & 0 \\
    Robust D & 98.96 & 2.30 & 104 & D & 18.00 & 92.21 & 9.84 & 99.91 & 99.91 \\
    Robust E & 99.14 & 1.93 & 6 & E & 62.68 & 91.86 & 6.12 & 99.21 & 99.21 \\
    \textbf{Robust A+B} & \textbf{99.74} & \textbf{0.56} & \textbf{84} & \textbf{A,B} & \textbf{99.68}  & \textbf{91.86} & \textbf{85.28} & \textbf{0} & \textbf{0} \\
    \textbf{Robust A+B+E} & \textbf{99.15} & \textbf{1.89} & \textbf{84} & \textbf{A,B,E} & \textbf{99.03} & \textbf{99.00} & \textbf{58.58} & \textbf{88.96} & \textbf{88.99} \\
    \hline
    \multicolumn{10}{l}{$*$ VRA numbers
    	are computed through the model property, not symbolic interval analysis.}
  \end{tabular}
  \caption{The verified robust accuracy (VRA) computed from 3,461 test PDF malware.
  The name of the monotonic classifier represents the number of trees in the model. For the other models,
  the name for the model corresponds to the property it is trained with. Although monotonic classifiers have higher VRA for insertion properties (B, D, E), Robust A+B and Robust A+B+E have strong VRA in both insertion and deletion properties, and therefore they are more robust against unrestricted attacks (Section~\ref{sec:Gradient Attackers} to Section~\ref{sec:Adaptive Evolutionary Attacker}).
  } 
  \label{tab:vra}
\end{table*}



\subsection{Bounded Arbitrary Attacker}
\label{sec:Any Attacker}

\textbf{Strongest Possible Bounded Attacker.}
The bounded arbitrary attacker has access to everything
(Table~\ref{tab:threat_model}).
The attacker can do anything to evade the classifier, under the restriction that
attacks are bounded by the robustness properties.

\subsubsection{Results}

We formally verify the robustness of the models
using symbolic interval analysis to obtain VRA, over all the 3,461 testing malicious PDFs (Table~\ref{tab:intervals}).
For example, 99\% VRA for property B means that, 99\% of 3,416 test PDFs
will always be classified as malicious,
for arbitrary insertion attacks restricted by one of the subtrees under the PDF root.
No matter how powerful the attacker is after knowing the defense, she will not have more than 1\% success rate.

Table~\ref{tab:vra} shows all the VRAs for the baseline models and verifiably robust models.
Our key observations are as follows.

\textbf{Baseline NN:} It has robustness for the deletion properties,
but not robust against insertion.

\textbf{Adversarially Robust Training:} All adversarially retrained models can increase the VRAs for deletion properties, except Adv Retrain B model.
Adv Retrain B model is trained with insertion at distance one, which shows conflict with the deletion properties
and decreased VRAs for property A and C, compared to the baseline NN.
Adv Retrain C achieves the highest VRAs for both property A and C.

\textbf{Ensemble Classifiers:} We conduct
the interval analysis according to the ensemble setup, described in Appendix~\ref{section:VRA for Ensemble Classifiers}.
Ensemble A+B has 97\% and 99\% VRAs for property A and B, respectively. However, the VRA
for property C is only 7\% and the VRA is zero for property D and E. On the other hand, Ensemble D
does not have VRA for any property, despite the ensemble setup. Since the
the base learner in Ensemble D needs to classify \emph{an arbitrary subtree} after certain deletion
and insertion operations, it is not enough to gain VRA by learning \emph{specific} subtree from the training dataset.

\textbf{Monotonic Classifiers:}
All the monotonic classifiers have insertion VRAs that are the same as the test accuracy, due to
the monotonic constraints enforced during the training time. Except the model with 10 learners,
all the models have over 99\% VRAs for properties B, D, and E.
We utilize the monotonic property of the models to find lower bound of deletion VRAs.
For property A, we verify the classifier's behavior on a malicious test PDF, if every possible mutated PDF with an arbitrary full subtree deletion is always classified correctly. Since the original malicious PDF features are larger, based on the monotonic property,
any partial deletion will also result in malicious classification for these PDFs. This gives us between 5.74\% and
8.78\% VRAs for the monotonic classifiers under property A. Similarly, by testing the lower bound of two
subtree deletion, we verify the monotonic classifiers to have 0 VRA for property C.


\textbf{Verifiably Robust Models:}
We can increase the VRA from as low as 0\% to as high as 99\%, maintaining high accuracy,
with under 0.6\% increase in FPR in properties A and B.

Training a model with one robustness property can make it obtain the same type of property
under a different distance. For example, Robust A model is trained
with property A (distance one), but it has also gained VRA in property C (distance two), that is
higher than the baseline NN model.

If we only train one type of property at a time, the other properties may be lost.
For example, Robust B and Robust D models both have decreased VRA in the deletion property,
compared to the baseline NN model. This indicates the conflicts between training for different tasks in general.



The strongest models according to the VRA evaluation are Ensemble A+B, monotonic classifiers,
Robust A+B, and Robust A+B+E.
While Adv Retrain A+B has slightly lower VRA than Adv Retrain C,
it is more robust against unrestricted gradient attack (Appendix~\ref{section:Unrestricted Gradient Attack Result}) since it is trained with more properties.
Robust A+B has slightly lower VRA in property B than the monotonic and ensemble baselines,
but it has 85.28\% VRA for property C. Robust A+B+E has gained VRA in all properties. Although the monotonic classifiers have higher VRA in insertion properties, since Robust A+B and Robust A+B+E have strong VRA in both insertion and deletion properties, they are more robust against unrestricted attacks than the monotonic classifiers, as shown by the results in the following sections.


\subsection{Gradient Attackers}
\label{sec:Gradient Attackers}



The gradient attacker has perfect knowledge, but the evasive feature vector
found by the attack may not correspond to a
real evasive PDF. We implement bounded and unbounded gradient attackers
to evaluate the ERA for all neural network models.

\subsubsection{Implementation}

\textbf{State-of-the-art Bounded Attacker.}
We implement the bounded gradient attacker for each robustness property. For example, for property B,
we first choose an arbitrary subtree from the PDF malware seed. Then, we take the gradient of the benign label with regard to the input feature, and increase the feature for the input index with the highest gradient value. We repeat this until either an evasion instance is found or the whole bounded region is inserted to be ones.
If any of the subtree choices succeeds, the attack can succeed
for the malware seed within the property. Similarly, we perform the bounded gradient attacks for the
other properties.


\textbf{State-of-the-art Unbounded Attacker.}
We implement the unbounded gradient attacker that performs arbitrary insertion and deletion guided by gradients, unrestricted by all robustness properties.
The attack stops when all evasive instances are found, or until 200,000 iterations.

\subsubsection{Results}

\begin{table}[t!]
  \begin{center}
  \small
    \begin{tabular}{lrrrrr}
     \hline
     & \multicolumn{5}{c}{Robustness Property (ERA, \%)} \\
      \textbf{Model} & \textbf{A} & \textbf{B} & \textbf{C} & \textbf{D} & \textbf{E} \\
      \hline
      Baseline NN &  98.51 & 0 & 88.44 & 0 & 0  \\
      Adv Retrain A+B & 99.8 & 84.6 & 91.42 & 87.3 & 94.7  \\
      Robust E & 67.1 & 99.27 & 19.15 & 99.27 & 99.27 \\
      Robust A+B & 99.77 & 99.97 & 91.04 & 0 & 0 \\
      Robust A+B+E & 99.56 & 99.91 & 90.66 & 99.21 & 99.21 \\
      \hline
      \end{tabular}
    \caption{ERA under bounded gradient attack. The corresponding VRAs
    in Table~\ref{tab:vra} are the lower bound of ERA values.}
    \label{tab:selected_era}
  \end{center}
\end{table}

\begin{table*}[ht]
\begin{minipage}[b]{0.25\textwidth}
\centering
\begin{tabular}{lr}
    \hline
      \textbf{Model} & \textbf{ERA (\%)} \\
      \hline
      Baseline NN &  0  \\
      Adv Retrain A+B & 0  \\
      Ensemble A+B & 0 \\
      Monotonic 100 & 48.8 \\
      Robust A+B & 0 \\
      Robust A+B+E & 50.8 \\
      \hline
      \end{tabular}
   \caption{ERA under reverse mimicry attack. Robust A+B+E is the most robust one.}
   \label{tab:reverse}
\end{minipage}\hfill
\begin{minipage}[b]{0.35\textwidth}
\includegraphics[scale=0.28]{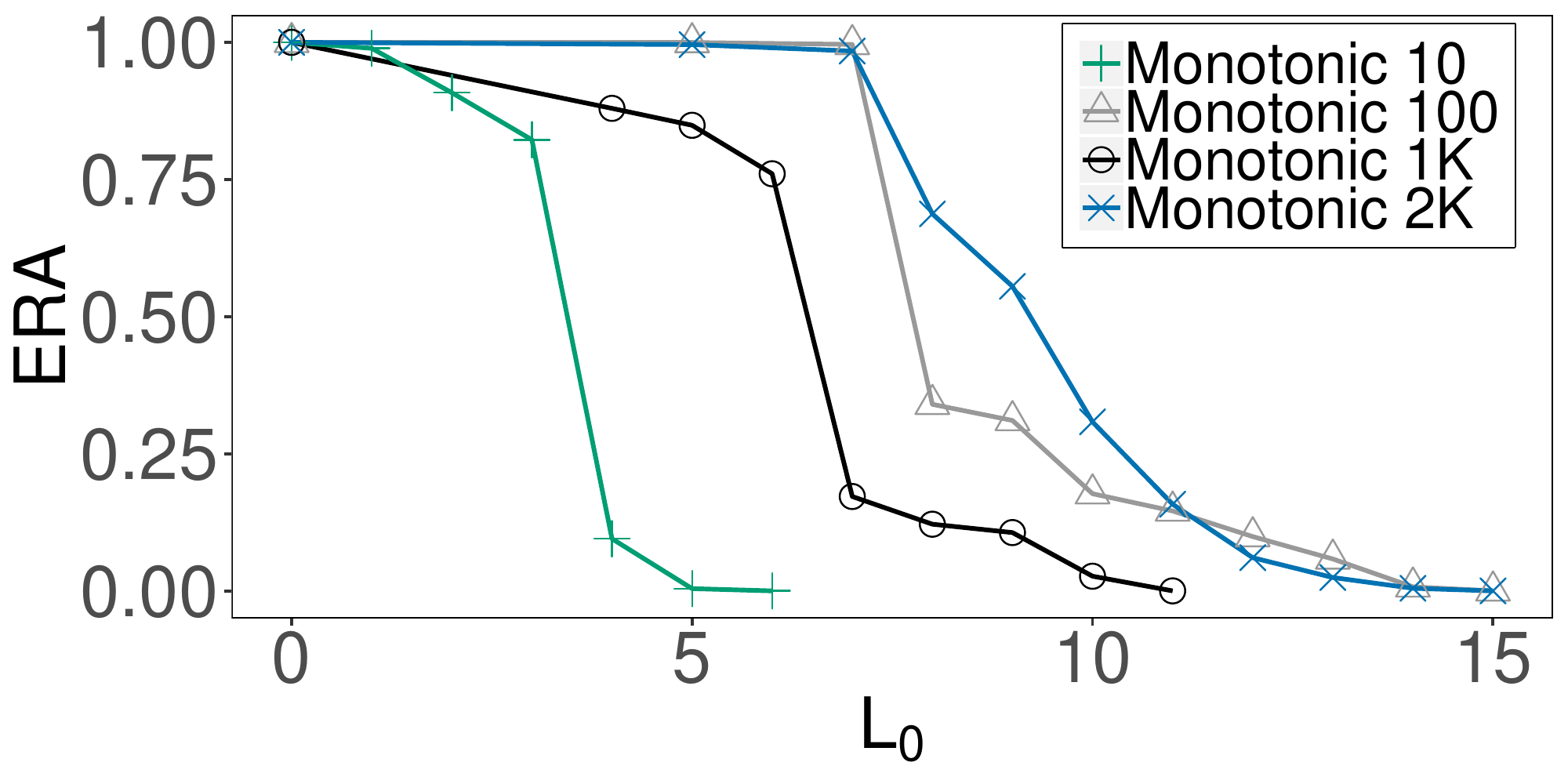}
\captionof{figure}{MILP attack: the monotonic classifiers can be evaded by up to 15 feature changes.}
\label{fig:era_monotonic}
\end{minipage}\hfill
\begin{minipage}[b]{0.35\textwidth}
\includegraphics[scale=0.28]{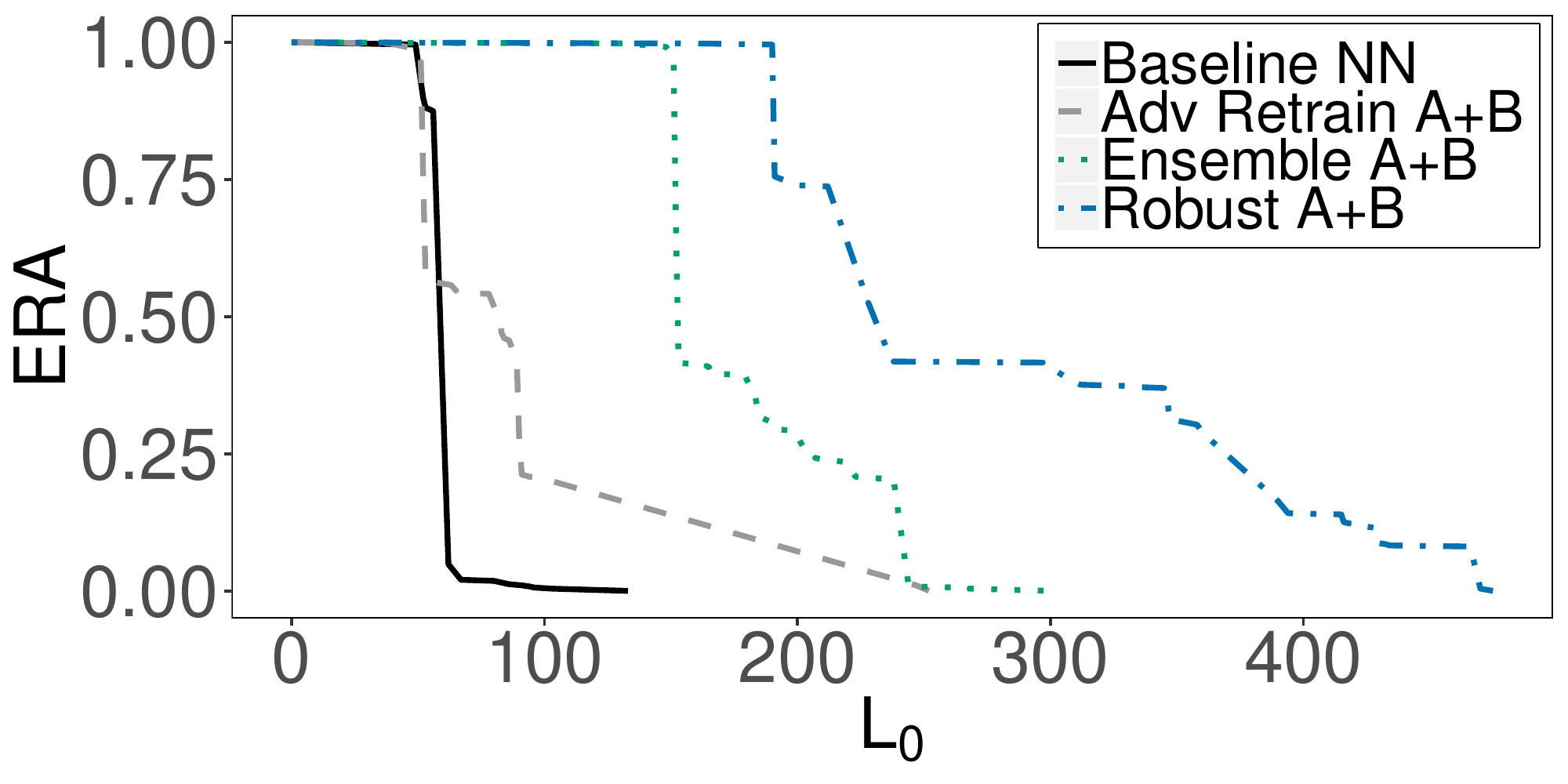}
\captionof{figure}{Enhanced evolutionary attack: Robust A+B requires up to 3.6 times larger $L_0$ distance to be evaded compared to the baselines.}
\label{fig:era_feat_nonadaptive}
\end{minipage}
\end{table*}

\textbf{Bounded Gradient Attack.}
We evaluate the ERA restricted by each robustness property.
Since VRA over-approximates possible attacks, VRA is the lower bound
for the ERA against any specific attack. We show the ERA for the most representative models in
Table~\ref{tab:selected_era}, with further details in Appendix~\ref{section:ERA under Bounded Gradient Attack}. All the ERA numbers are higher
than the corresponding VRA in Table~\ref{tab:vra}.
For the Adv Retrain A+B model, the gap between
VRA and ERA is quite large. While it has 0 VRA for property B and D (Table~\ref{tab:vra}), the ERA values under
bounded gradient attack are 84.6\% and 87.3\% respectively.
One interesting observation is that the ERA of property E is higher than
property D, which is also higher than property B. It shows that the greedy algorithm (gradient descent) is
not effective at evading this model, since the attack could have used the solution from
property B to reduce the ERA for the other two properties to at least 84.6\%.
Although Adv Retrain A+B has a higher ERA in property D and E against the gradient attack
than the Robust A+B model, both models have 0 VRA in these properties.
Since there always exist stronger attacks that can
reduce the ERA~\cite{wang2019enhancing}, VRA provides
a stronger robustness guarantee than ERA.

\textbf{Unbounded Gradient Attack.}
Our verifiably robust model can improve the ERA
against the unrestricted gradient attack by 7\% up to 200,000 iterations.
Table~\ref{tab:gradient_era_acc_comparison} shows the attack results
on five representative models. The unrestricted gradient attack can
reduce the ERA for three models to 0 given enough allowable $L_0$ distance
in feature changes.
Baseline NN is the first to reach 0 ERA at $L_0 = 19$, whereas
Robust A+B+E requires the largest $L_0$ distance (68) to reach 0 ERA.
For the other three models, Robust A+B is the most robust one. It
maintains 7.38\% ERA
even after 200,000 attack iterations. The ERA converges
for Adv Retrain A+B, Robust D and Robust A+B models
against the unrestricted gradient attack (Appendix~\ref{section:Convergence}).
Further details about the attack can be found in Appendix~\ref{section:Unrestricted Gradient Attack Result}.


We convert the evasive feature vectors to real PDFs.
Given each feature index change, we either delete the corresponding PDF object,
or insert the object
with minimal number of children in the benign training dataset.
Inserting object with minimal children makes the features from constructed PDF close to the evasive features.
On average, the ERA of models against the real evasive PDF malware is
94.25\%, much higher than 0.62\% ERA against evasive feature vectors, since
unrestricted gradient attack often breaks the PDF semantics
(Appendix~\ref{section:Real PDFs from Unrestricted Gradient Attack}).
Due to the inherent limitation of feature-space attacks, we also evaluate robustness of the models against realizable attacks from Section~\ref{sec:Enhanced Evolutionary Attacker} to Section~\ref{sec:Adaptive Evolutionary Attacker}.

\begin{table}[t!]
  \begin{center}
  \small
    \begin{tabular}{lrr}
    \hline
      & \textbf{$L_0$ for} & \textbf{ERA (\%) at 200K} \\
      \textbf{Model} & \textbf{ERA=0} & \textbf{attack iterations} \\
      \hline
      Baseline NN & 19 & 0  \\
      Adv Retrain A+B & N/A & 0.32 \\
      Robust A & 36 & 0 \\
      Robust D & N/A & 0.03 \\
      Robust A+B & N/A & 7.38 \\
      Robust A+B+E & 68 & 0 \\
      \hline
      \end{tabular}
      \caption{Robust A+B model maintains 7\% higher ERA against the unrestricted gradient attack than the other five models.
      \label{tab:gradient_era_acc_comparison}}
  \end{center}
\end{table}

\subsection{MILP Attacker}

\textbf{State-of-the-art Unbounded Attacker.}
The Mixed Integer Linear Program (MILP) attacker is the unbounded whitebox attacker
for the GBDT monotonic classifiers, proposed by Kantchelian et al.~\cite{kantchelian2016evasion}.
The attack
formulates the evasion problem as a Mixed Integer Linear Program. The variables in this program
are predicates and leaves in all the decision trees. We set the objective of the linear program to minimize
the $L_0$ distance between the seed malware feature vector and the variant vector.
The constraints to solve the linear program include
model misclassification, consistency among leaves and predicates, and several variables being integer.
The solution to the MILP represents the malware feature vector manipulation.
We use the re-implementation from~\cite{chen2019robust} to conduct the MILP attack.

\subsubsection{Result}

The MILP attack can succeed for all the 3416 test malicious PDFs against all four
monotonic classifiers. We plot the ERA values with different
$L_0$ distance (number of feature changes) in Figure~\ref{fig:era_monotonic}.
The Monotonic 10 model is the weakest among them. With only 2 feature deletion,
10\% of PDFs can be evaded, e.g., deleting
\texttt{/Root/Names/JavaScript/Names} and \texttt{/Root/Names/JavaScript/Names/JS/Filter}.
Everything can be evaded by up to six feature changes for the 10 learner model.
Using 100 learners can increase the $L_0$ distance for evasion.
However, using more learners does not increase the robustness after 100 learners.
All the monotonic classifiers can be evaded by up to 15 feature changes.
In comparison, when $L_0=15$, the ERA for Robust A+B is 10.54\%.
Under different whitebox attacks, Robust A+B is more robust than the monotonic classifiers.

After converting the evasive feature vectors to real PDFs, none of them are still malicious,
since the MILP attack deletes the exploit (Appendix~\ref{section:Real PDFs from Unrestricted Gradient Attack}).
Next, we will evaluate the strongest models against unrestricted black box attacks that ensure
the maliciousness of evasive PDF variants.


\subsection{Enhanced Evolutionary Attacker}
\label{sec:Enhanced Evolutionary Attacker}

\textbf{State-of-the-art Unbounded Attacker.}
The enhanced evolutionary attacker has black-box oracle access to the model, including the classification label and scores, and she is not bounded by the robustness properties.
The attack is based on the genetic evolution algorithm~\cite{xu2016automatically}.

\subsubsection{Implementation}
\label{section:evademl_implementation}

The genetic evolution attack evades the model prediction function by
mutating the PDF malware, using random deletion, insertion, and replacement,
guided by a fitness function.
We implemented two strategies to enhance the evolutionary attack,
with details and the experiment set up in
Appendix~\ref{section:Genetic Evolution Attack}.



\subsubsection{Results}
\label{sec:nonadaptive_results}

Within the 500 PDF malware seeds that exhibit network behavior from previous work~\cite{xu2016automatically}, 
we can detect 495 PDFs with signatures using our cuckoo sandbox.
All the PDF malware seeds belong to the testing PDF set in Table~\ref{tab:dataset}.
By round robin, we go through the list of randomly scheduled PDFs by rounds of attacks, until all
of them are evaded.



We run the attack on the best baseline and robust models: Baseline NN, Adv Retrain A+B,
Ensemble A+B, Monotonic Classifiers, Robust A+B, and Robust A+B+E.
For four models without property E, the attack has succeeded in generating evasive variants for
all PDF seeds. It takes between three days to two weeks to evade each model.
The attack is not effective against monotonic classifier and Robust A+B+E model.
Although the attack can identify that deletion is preferred to evade the models, sometimes it deletes the exploit.
We design adaptive evolutionary attacks to evade these models in Section~\ref{sec:Adaptive Evolutionary Attacker}.



\textbf{$L_0$ distance.}
The enhanced evolutionary attack needs up to 3.6 times larger $L_0$ distance,
and 21 times more mutations (Appendix~\ref{section:Trace Length of Evolutionary Attacks}) to evade our robust model than the baselines.
We plot the ERA for different models under various $L_0$ distances to generate evasive PDF variants in Figure~\ref{fig:era_feat_nonadaptive}.
For hidost features, the $L_0$ distance also means the number of feature changes.
To evade the baseline NN model, at least 49 features need to be changed.
The ERA of the model quickly drops to zero at 133 features. The Adv Retrain A+B and Ensemble A+B both require more changes to
be fully evaded, up to 252 and 300 respectively. Compared to these baselines, our Robust A+B model
needs the most number of feature changes (475) to be evaded, 3.6 times of that against the Baseline NN.
The smallest $L_0$ distances
to generate one evasive PDF malware variant are 49, 39, 134, and 159 for Baseline NN, Adv Retrain A+B,
Ensemble A+B, and Robust A+B, respectively.

\subsection{Reverse Mimicry Attacker}

\textbf{State-of-the-art Unbounded Attacker.}
The reverse mimicry attacker injects malicious payload into a benign PDF,
which is outside of all five robustness properties. We have proposed robustness properties
for malicious PDFs, not benign ones.
The attacker uses the same strategy for all models, and thus she does not need to know model internals or
the defenses.

\subsubsection{Implementation}

We implement our own reverse mimicry attack, similar to the JSinject~\cite{maiorca2013looking}.
We use peepdf~\cite{peepdf} static analyzer to identify the suspicious objects in the PDF malware seeds, and then inject these objects to a benign PDF. We inject different malicious payload
into a benign file, whereas the JSinject attack injects the same JavaScript code into different benign PDFs.
Within the PDF malware seeds, 250 of them retained maliciousness according to the cuckoo oracle.
Some payload are no longer malicious because there can be object dependencies within the malware not identified by the static analyzer. We test whether the models can detect the 250 PDFs are malicious.

\subsubsection{Results}

We measure ERA as the percentage of correctly classified PDFs for the strongest models
against whitebox attacks in Table~\ref{tab:reverse}. Since this is outside all five robustness properties, the attack
can defeat most verifiably robust models and baseline models, except the monotonic classifier and Robust A+B+E
models. The monotonic classifier has the monotonic constraint enforced for the benign PDFs, whereas
we only trained property E for malicious PDFs for our Robust A+B+E model. However, we still achieve
2\% higher ERA than the Monotonic 100 model against the reverse mimicry attack. This shows
that verifiably robust training can generalize outside the trained robustness properties.
Since training property E incurs higher FPR than properties with smaller subtree distances,
we plan to experiment with training insertion property with small distance for benign samples as future work.

\subsection{Adaptive Evolutionary Attacker}
\label{sec:Adaptive Evolutionary Attacker}

\textbf{New Adaptive Unbounded Attacker.}
The adaptive evolutionary attacker has the same level of black-box access as
the enhanced evolutionary attacker (Section~\ref{sec:Enhanced Evolutionary Attacker}).
She is not bounded by the robustness properties and knows about the defense.

\subsubsection{Implementation}

To evade the three strongest models: the monotonic classifier, Robust A+B, and Robust A+B+E,
we design three versions of the adaptive attacks as following.


\textbf{Move Exploit Attack.}
The monotonic property forces the attacker to delete objects from the malware, but deletion could remove the exploit.
Therefore, we implement a new mutation to move the exploit around to different
trigger points in the PDF (Appendix~\ref{section:Move Exploit}).
This attack combines
the move exploit mutation with deletion to evade the monotonic classifier.
Note that the move exploit mutation is not effective against Robust A+B, since it is
covered by the insertion and deletion properties.


\textbf{Scatter Attack.}
To evade Robust A+B, we insert and delete more objects
under different subtrees. We keep track of past insertion and deletion operations separately,
and prioritize new insertion and deletion operations to target a different subtree.




\textbf{Move and Scatter Combination Attack.}
To evade the Robust A+B+E model, we combine the move exploit attack and the scatter attack,
to target all the properties of the model.

\subsubsection{Results}

\begin{figure}[t]
    \centering
    \scalebox{0.3}{\includegraphics{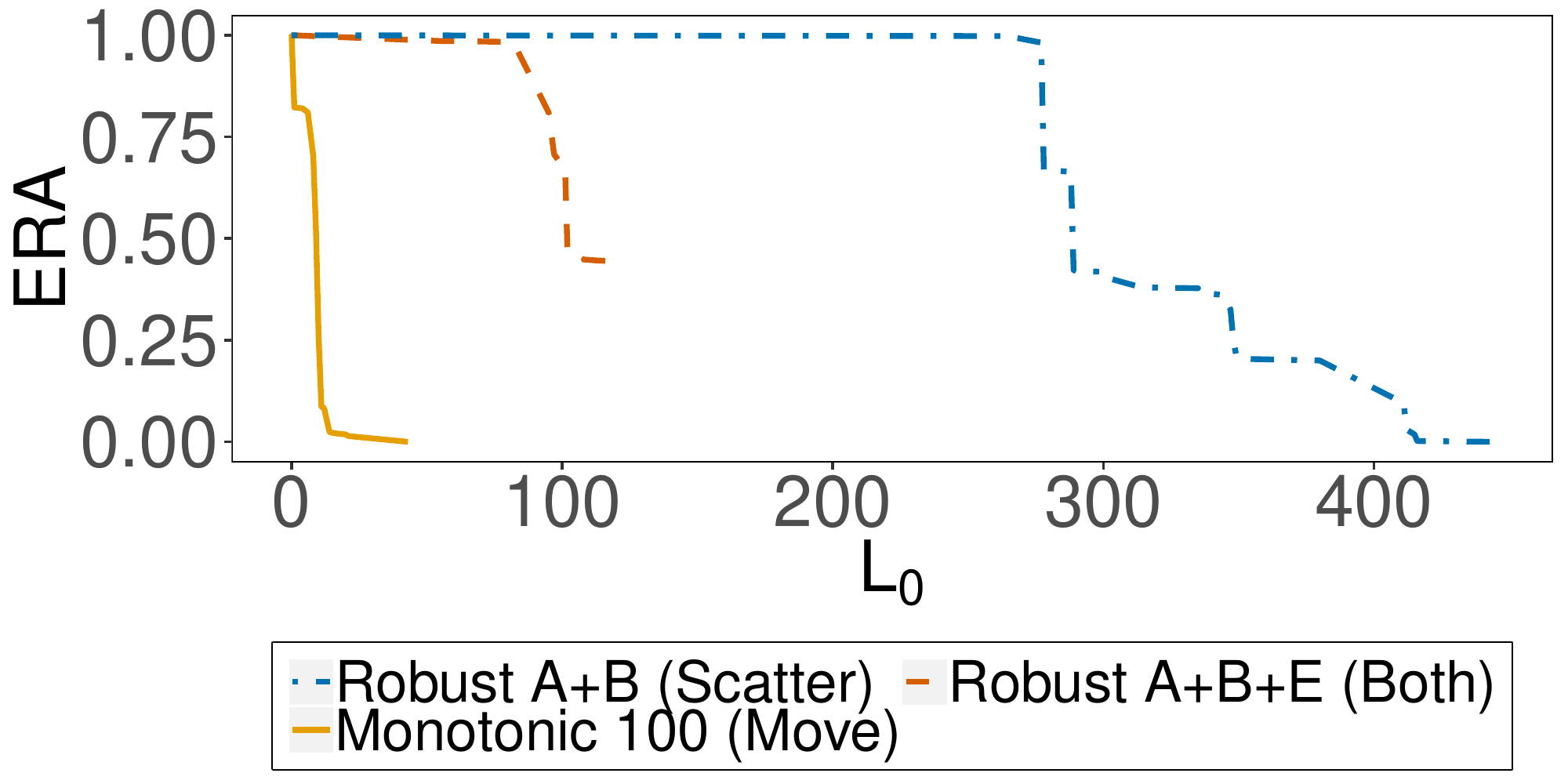}}
    \caption{The decrease of robustness in ERA against adaptive evolutionary attacks as the $L_0$ distance increases.
    }
   \label{fig:era_feat_adaptive}
\end{figure}


The adaptive attacks need 10 times larger $L_0$ distance (Figure~\ref{fig:era_feat_adaptive}), and 3.7 times more mutations (Appendix~\ref{section:Trace Length of Evolutionary Attacks}) to evade our model than the monotonic classifier.
Figure~\ref{fig:era_feat_adaptive} shows the $L_0$ distance to evade the three models:
Monotonic 100, Robust A+B,
and Robust A+B+E. The move exploit attack is very
effective against the Monotonic 100 model. The ERA of Monotonic 100 quickly drops
to zero at $L_0 = 10$.
The scatter attack
can reduce the mutation trace length to evade Robust A+B
compared to the nonadaptive version. However, the median $L_0$ distance
has increased from 228 (Figure~\ref{fig:era_feat_nonadaptive}) to 289 (Figure~\ref{fig:era_feat_adaptive}).
The minimal $L_0$ distances to generate one evasive PDF malware for the Monotonic 100 and Robust A+B
are 1 and 263 respectively.
Lastly, the move and scatter combination attack can reduce the ERA of Robust A+C+E
to 44\% after running for three weeks. The attack is stuck at premature convergence and needs
additional improvements to fully evade the model.

\section{Discussion}

\textbf{Generalization.}
In the arms race against malware detection and evasion, there has been no verifiably 
robust solution to the detection problem.
By setting bounds on attackers' actions, we can provide verifiable robustness
properties in PDF malware classifiers.
We further show that such robust training can also increase the bar
for state-of-the-art unbounded attackers.
Since we specify robustness properties related to the PDF syntax,
they can be generalized to different features, datasets, and models. Our method
can be complementary to other defenses such as feature reduction.
We plan to explore all these issues regarding the generalization of our methodology in our future work.

\textbf{Scalability.}
Verifiably robust training using symbolic interval analysis is faster than existing sound approximation methods, achieving state-of-the-art tight bounds.
Many techniques can scale the training to larger neural networks with hundreds of thousands of hidden units, and larger datasets such as ImageNet-200~\cite{gowal2018effectiveness,wang2018mixtrain,shiqi2018efficient,wong2018scaling}. We plan to explore the tradeoffs between scalability and performance (e.g., accuracy, robustness, and false positive rate) of the trained network.

\textbf{Diverse Robustness Properties.}
The robustness properties for insertion and deletion can be used as building
blocks to construct stronger properties.
Training combinations of properties can make the evasion task even harder for the attacker.
In addition, we plan to
train verifiable robustness properties for benign PDFs, to defend against another type
of evasion search that starts from a benign PDF seed. Exploring the tradeoffs among
learning multiple robustness properties and overhead of training
will be an interesting direction for future work.





%
%
%

\section{Related Work}

Existing defenses in increasing the robustness of malware classifiers mainly focus on
using feature reduction and adversarially robust retraining. Researchers have employed methods including mutual information~\cite{grosse2016adversarial}, expert domain knowlege~\cite{incer2018adversarially},
information from cuckoo sandbox~\cite{tong2019improving} to remove features
unrelated to maliciousness.
However, previous adversarial retraining results show severe drop in accuracy~\cite{incer2018adversarially}, and increase
in false positive rate~\cite{advtrain_pdf_slides, grosse2016adversarial}.



Incer et al.~\cite{incer2018adversarially} enforced the monotonicity property to make the malware
classifier robust
against attacks that increase feature values. Thus, attackers have to conduct more expensive
feature manipulation that might remove the malicious functionality. In comparison, we train
robustness properties not only for insertion, but also for deletion, since deletion operations are
often not costly to the attacker~\cite{xu2016automatically}.

Our method can increase the feature distance and mutation trace length as cost for the attacker to evade the model.
Existing works have discussed cost for the attackers to manipulate features~\cite{lowd2005adversarial},
to increase suspiciousness~\cite{chen2017practical}, and to solve the combinatorial optimization problem~\cite{dai2018adversarial}.
On the other hand, several work have explored the cost for the defender~\cite{zhang2018cost, dreossi2018semantic}.
Dreossi et al.~\cite{dreossi2018semantic} argued that only some adversarial examples
cause the overall control system to make catastrophic decision.
Zhang et al.~\cite{zhang2018cost} integrated the defender's cost with Wong et al.'s verifiably robust training
method~\cite{wong2018provable}.

\section{Conclusion}

We are the first to train verifiable robustness properties for PDF malware classifier.
We proposed a new distance metric in the PDF tree structure to bound
robustness properties. Our best model achieved 99.68\% and 85.28\% verified robust accuracy (VRA)
for the insertion and deletion properties, while maintaining \bestmodelacc{} accuracy and \bestmodelfpr{}
false positive rate.
Our results showed that training security classifiers with verifiable robustness properties
is a promising direction to increase the bar for unrestricted attackers. 


\section*{Acknowledgements}

We thank our shepherd Nicolas Papernot and the anonymous reviewers for their constructive and valuable feedback. This work is sponsored in part by NSF grants CNS-18-42456, CNS-18-01426, CNS-16-17670, CNS-16-18771, CCF-16-19123, CCF-18-22965, CNS-19-46068; ONR grant N00014-17-1-2010; an ARL Young Investigator (YIP) award; a NSF CAREER award; a Google Faculty Fellowship; a Capital One Research Grant; and a J.P. Morgan Faculty Award. Any opinions, findings, conclusions, or recommendations expressed herein are those of the authors, and do not necessarily reflect those of the US Government, ONR, ARL, NSF, Google, Capital One or J.P. Morgan.

\small
\bibliographystyle{abbrv}
\bibliography{ref}

\appendix
\normalsize
\section{Appendix}

\subsection{VRA for Ensemble Classifiers}
\label{section:VRA for Ensemble Classifiers}
\subsubsection{Ensemble A+B VRA}

\paragraph{Property A.} A test PDF is verified to be safe within property A, if \emph{all} the
possible subtree deletion with distance one is safe. Therefore, for each interval
representing one subtree deletion, we require that \emph{any} of the corresponding
two subtree deletion is classified as malicious.

\paragraph{Property B.} Property B is the provable robustness property of Ensemble A+B.
If any mutated PDF is generated by inserting one arbitrary subtree to a malicious PDF,
it has the same classification result as the malicious PDF seed. Therefore,
we use the test accuracy of malicious PDFs as the VRA for property B.

\paragraph{Property C.} A test PDF is verified to be safe within property C, if \emph{all} the
possible subtree deletion with distance two is safe. Therefore, for each interval
representing two subtree deletion, we require that \emph{any} of the corresponding
three subtree deletion is classified as malicious.

\paragraph{Property D.} A test PDF is verified to be safe within property D, if \emph{all}
the possible subtree insertion at distance 41 is safe. Therefore, we test
whether \emph{any} interval representing 40 subtree insertion on a malicious test PDF
can be classified as malicious.

\paragraph{Property E.} A test PDF is verified to be safe within property E, if \emph{all}
the possible subtree insertion in the entire feature space is safe. Therefore, we test
whether \emph{any} interval representing all-but-one (41) subtree insertion on a malicious test PDF
can be classified as malicious.

\begin{table*}[ht!]
  \centering
  \small
  \begin{tabular}{lrrr | rrrrr}
    \hline
    & & & & \multicolumn{5}{c}{Estimated Robust Accuracy (VRA, \%) against Bounded Gradient Attacker} \\
    & & & & \multicolumn{1}{c}{Property A} & \multicolumn{1}{c}{Property B} & \multicolumn{1}{c}{Property C} & \multicolumn{1}{c}{Property D} & Property E \\
   & \textbf{Precision} & \textbf{Recall} & \textbf{Trained} & \multicolumn{1}{c}{Distance: 1} & \multicolumn{1}{c}{Distance: 1} & \multicolumn{1}{c}{Distance: 2} & \multicolumn{1}{c}{Distance: 41} & \multicolumn{1}{c}{Distance: 42} \\
    \textbf{Model} & \textbf{(\%)} & \textbf{(\%)} & \textbf{Prop.} & \textbf{Subtree Del.} & \textbf{Subtree Ins.} & \textbf{Subtree Del.} & \textbf{Subtree Ins.}  & \textbf{Subtree Ins.}\\
    \hline
    \hline
    Baseline NN & 99.94 & 99.97 & None & 98.51 & 0 & 88.44 & 0 & 0 \\
    \hline
    \hline
    Adv Retrain A & 99.94 & 99.97 & A & 99.53 & 0 & 88.2 & 0 & 0 \\
    Adv Retrain B & 99.94 & 99.97 & B & 89.26 & 9.57 & 60.91 & 14.93 & 14.58 \\
    Adv Retrain C & 99.91 & 99.97 & C & 99.47 & 0 & 91.43 & 0 & 0 \\
    Adv Retrain D & 99.91 & 99.97 & D & 97.51 & 0 & 61.18 & 0 & 0 \\
    Adv Retrain A+B & 99.88 & 99.97 & A,B & 99.8 & 84.6 & 91.42 & 87.3 & 94.7 \\
    \hline
    \hline
    Ensemble A+B* & 99.97 & 99.97 & A,B & 99.5 & 0 & 20.19 & 0 & 0 \\
    Ensemble D* & 99.94 & 99.97 & D & 99.24 & 0 & 88.17 & 0 & 0 \\
    \hline
    \hline
    Robust A & 99.74 & 99.97 & A & 99.85 & 0 & 99.53 & 0 & 0 \\
    Robust B & 99.54 & 99.97 & B & 50.06 & 99.97 & 27.61 & 0 & 0 \\
    Robust C & 99.19 & 100 & C & 99.94 & 0 & 99.82 & 0 & 0 \\
    Robust D & 98.23 & 99.94 & D & 66.28 & 99.94 & 22.34 & 99.91 & 99.91 \\
    Robust E & 98.51 & 99.97 & E & 67.1 & 99.27 & 19.15 & 99.27 & 99.27 \\
    Robust A+B &  99.57 & 99.97 & A,B & 99.77 & 99.97 & 91.04 & 0 & 0 \\
    Robust A+B+E & 98.54 & 99.97 & A,B,E & 99.56 & 99.91 & 90.66 & 99.21 & 99.21 \\
    \hline
  \end{tabular}
  \caption{The estimated robust accuracy (ERA) against bounded gradient attacker,
  computed from 3,461 testing PDF malware,
  over five robustness properties.
  *We run bounded gradient attack against the base learner of ensemble models.
  } 
  \label{tab:era}
\end{table*}

\subsubsection{Ensemble D VRA}

\paragraph{Property A and C.}
A test PDF is verified to be safe for a deletion property, if \emph{any} subtree
after some deletion is classified as malicious. Therefore, for each test PDF,
we check whether any interval representing the lower bound of all zeros and
the upper bound of the original subtree can be classified as malicious.

\paragraph{Property B, D and E.}
A test PDF is verified to be safe for a insertion property, if \emph{any} subtree
\emph{after some insertion} is classified as malicious. There are two categories.
If the inserted subtree does not exist, the interval is from all zeros and
all ones for that subtree. If the inserted subtree already exists, the interval
bound is from the original subtree features to all ones. We check if any of
these intervals can be classified as malicious for all possible insertions.

\subsection{ERA under Bounded Gradient Attack}
\label{section:ERA under Bounded Gradient Attack}

Table~\ref{tab:era} shows precision, recall of the models on the left side,
and the ERA under gradient attacks bounded by robustness properties
on the right side. All verifiably robust models maintain high precision and recall.
The ERA values of the models are higher than the corresponding VRA values in Table~\ref{tab:vra}.

%

\subsection{Unrestricted Gradient Attack Result}
\label{section:Unrestricted Gradient Attack Result}

\subsubsection{ERA}

\begin{figure}[th!]
    \centering
    \scalebox{0.3}{\includegraphics{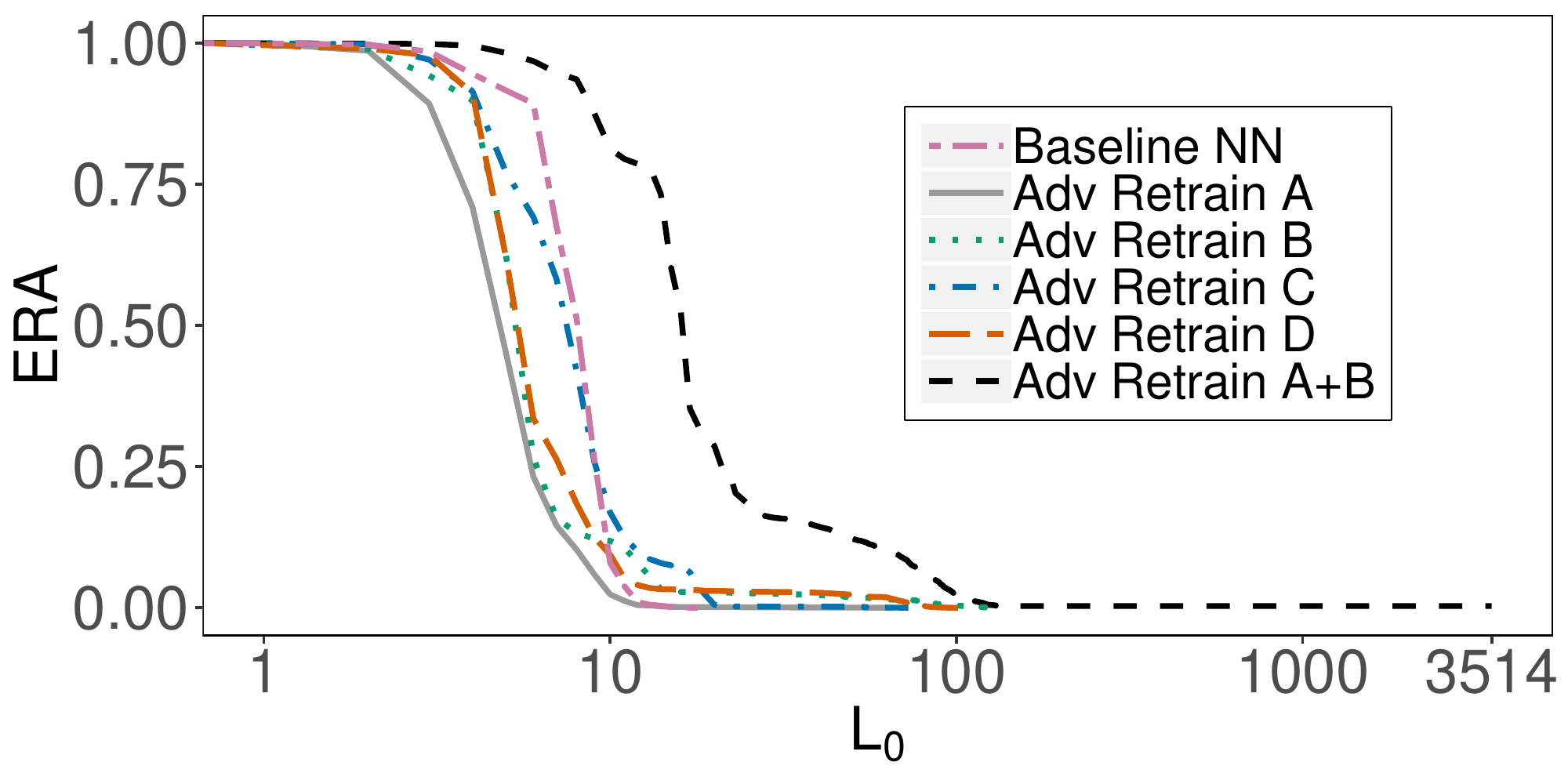}}
    \caption{
    Unrestricted gradient attack against baseline models.
    }
   \label{fig:gradient_era_acc_baseline}
\end{figure}

\begin{figure}[th!]
    \centering
    \scalebox{0.3}{\includegraphics{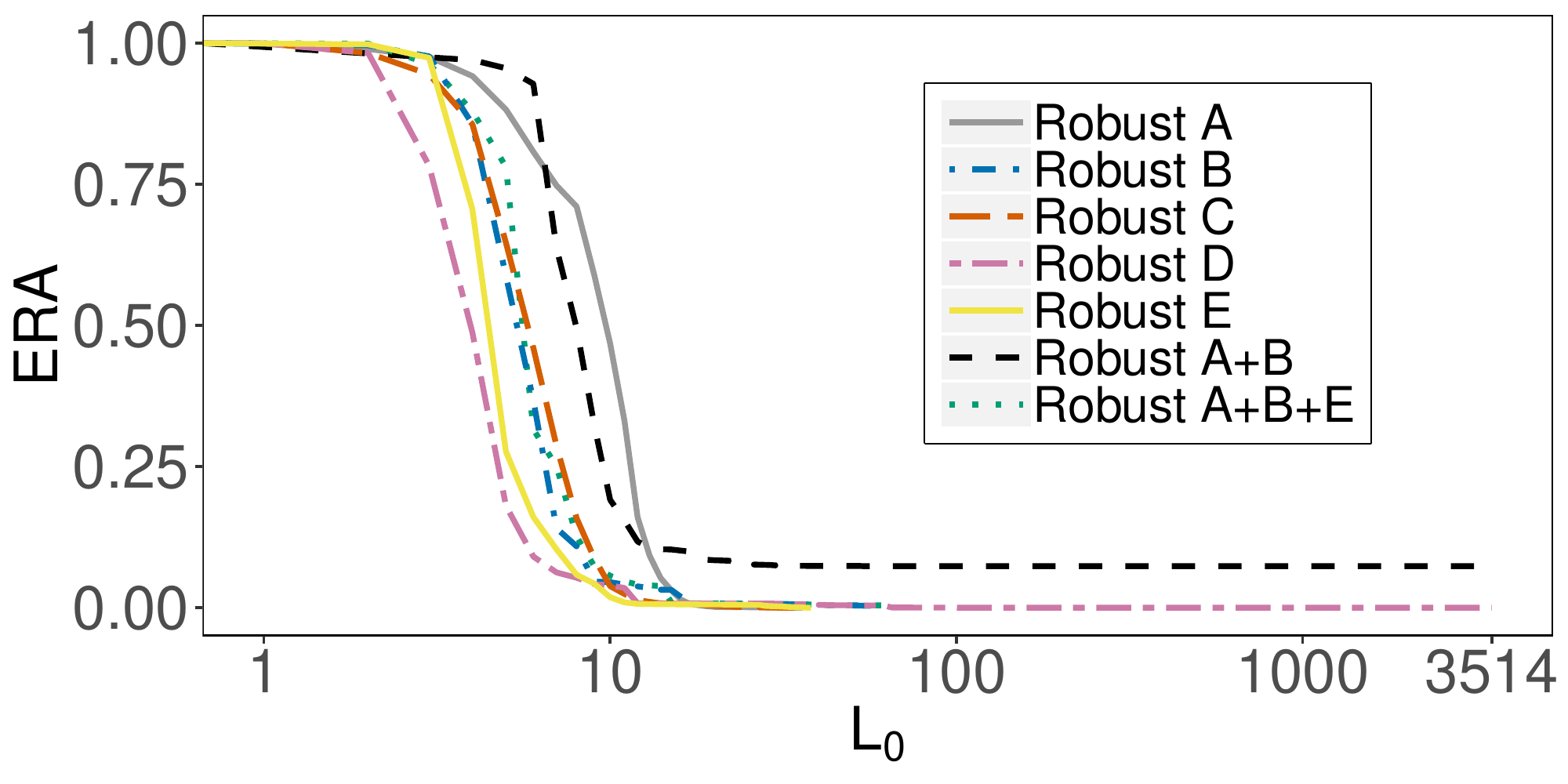}}
    \caption{
    Unrestricted gradient attack against our verifiably robust models.
    }
   \label{fig:gradient_era_acc_robust}
\end{figure}
\vspace{10pt}

Figure~\ref{fig:gradient_era_acc_baseline} shows the ERA of the Baseline NN and adversarially retrained models against unrestricted gradient attack.
Most adversarially retrained models perform similar to the Baseline NN. Adv Retrain A+B
is most robust among them according to the ERA curve. The ERA drops more slowly as
the $L_0$ distance increases compared to the other models.

Figure~\ref{fig:gradient_era_acc_robust} shows the ERA of verifiably robust models
against unrestricted gradient attack. Robust A+B performs the best among them, maintaining
7.38\% ERA after 200,000 attack iterations.

\subsubsection{Convergence}
\label{section:Convergence}

\begin{figure}[th!]
    \centering
    \scalebox{0.3}{\includegraphics{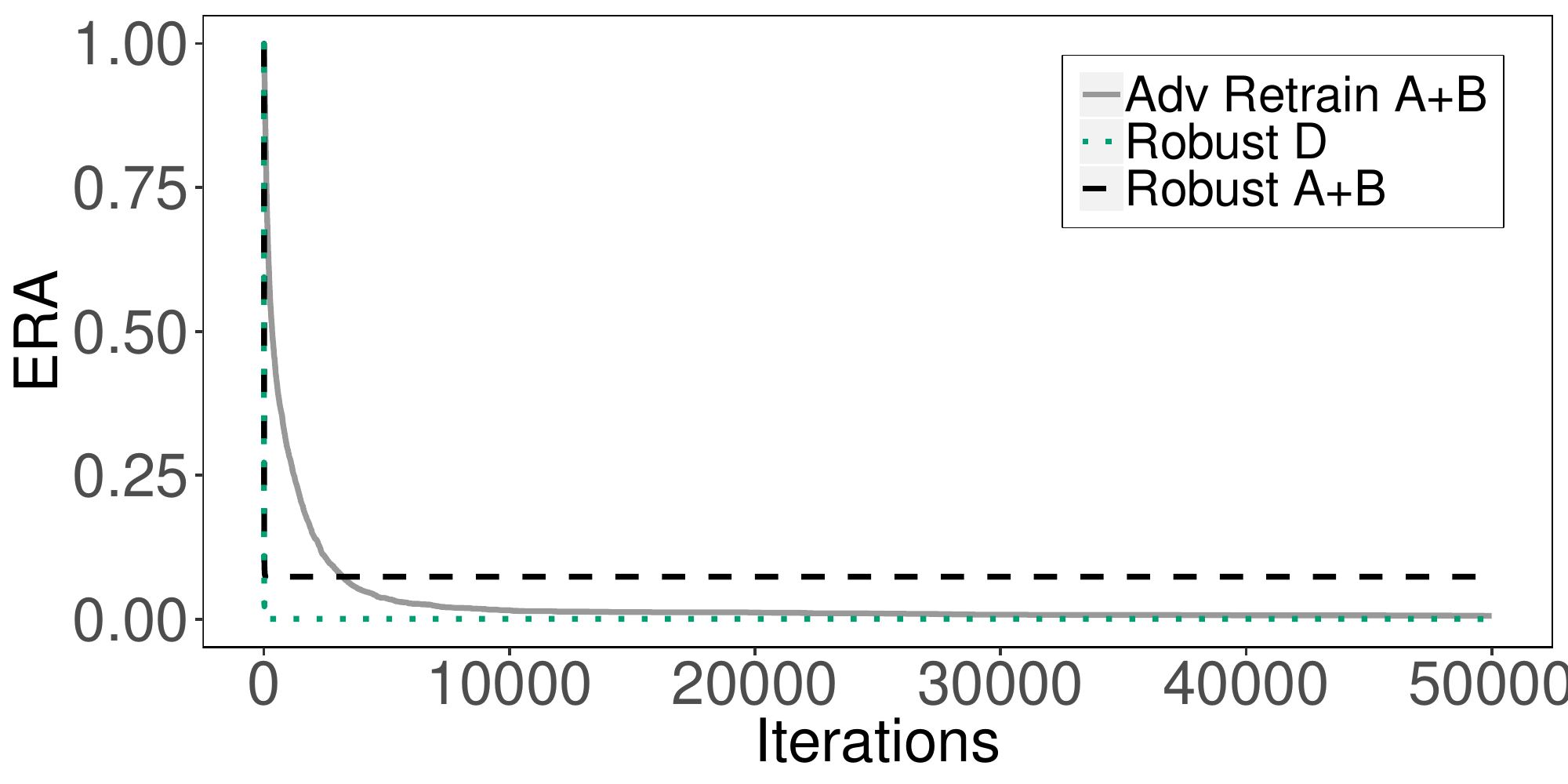}}
    \caption{
    The ERA of three models converges against the unrestricted gradient attack.
    }
   \label{fig:un_gradient_era_acc_converge}
\end{figure}
\vspace{10pt}

We run the unrestricted gradient attack for 200,000 iterations, and plot the ERA for the first 50,000 iterations.
Figure~\ref{fig:un_gradient_era_acc_converge} shows that the unrestricted gradient attack converges for
Adv Retrain A+B, Robust D, and Robust A+B models. The ERA of Robust D model stays the same after 49,128 attack iterations, and the ERA of Robust D A+B model stays the same after 60 attack iterations.
For Adv Retrain A+B, the ERA only decreases very slowly between 30,000 and 200,000 iterations.

\subsection{Real PDFs from Unrestricted Gradient Attack}
\label{section:Real PDFs from Unrestricted Gradient Attack}

Within the 3,416 evasive PDF vectors, we convert 495 of them back to
real PDFs, i.e., those with network signature that can be detected by the
cuckoo sandbox. Then, we measure the ERA for features extracted
from the real PDFs as the 3rd column in Table~\ref{tab:realpdf_era}.
Furthermore, we check how many of these PDFs are still malicious
using the cuckoo sandbox. Then we measure the ERA against the end-to-end
attack that generates malicious PDFs, as the last column in Table~\ref{tab:realpdf_era}.

\begin{table}[t!]
  \begin{center}
  \small
    \begin{tabular}{lrrr}
     \hline
     & \multicolumn{3}{c}{ERA(\%)} \\
      \textbf{Model} & \textbf{Feature} & \textbf{PDFs} & \textbf{Malicious PDFs} \\
      \hline
      Baseline NN &  0 & 28.48 & 98.78  \\
      \hline
      Adv Retrain A &  0 & 7.88 & 92.93  \\
      Adv Retrain B &  0 & 27.68 & 100  \\
      Adv Retrain C &  0 & 8.48 & 92.53  \\
      Adv Retrain D &  0 & 22.83 & 100  \\
      Adv Retrain A+B & 0.6 & 39.39 & 100  \\
      \hline
      Robust A &   0 & 2.02 & 90.71 \\
      Robust B &   0 & 88.28 & 100 \\
      Robust C &   0 & 5.86 & 74.55 \\
      Robust D &   0.03 & 58.38 & 99.6 \\
      Robust E &   0 & 28.28 & 88.28 \\
      Robust A+B & 7.38 & 99.8 & 100 \\
      Robust A+B+E & 0 & 40.61 & 87.88 \\
      \hline
      NN Models Average & 0.62 & 35.23 & 94.25 \\
      \hline
      \hline
      Monotonic 10 & 0 & 32.73 & 100 \\
      Monotonic 100 & 0 & 0 & 100 \\
      Monotonic 1K & 0 & 0 & 100 \\
      Monotonic 2K & 0 & 0 & 100 \\
      \hline
      Monotonic Average & 0 & 8.18 & 100\\
      \hline
      \end{tabular}
    \caption{ERA according to feature vectors, corresponding PDFs, and corresponding malicious PDFs.}
    \label{tab:realpdf_era}
  \end{center}
\end{table}

Although the neural network models have an average of 0.62\% ERA against evasive
feature vectors, that increases to 35.23\% if we enforce that they classify the
corresponding PDF files. The average ERA further increases to 94.25\%
if we require that the generated PDFs are malicious.

For monotonic classifiers, the average ERA against evasive feature vectors is 0\%,
which increases to 100\% if we require the corresponding evasive PDF to be malicious.
This is because the MILP solver always finds the action that deletes the exploit
to evade the monotonic property.

\subsection{Genetic Evolution Attack}
\label{section:Genetic Evolution Attack}

\subsubsection{Fitness Function}
To construct the fitness function for neural network, we take the output of softmax as
the classification scores for malicious and benign classes, and compute $\log{(benign)} - \log{(malicious)}$.
This helps prioritize PDF variants
with very small prediction changes in the floating point number.
When the fitness score reaches zero,
the attack succeeds.


\subsubsection{Two Improvement Strategies}
First, we improve the effectiveness of insertion and replacement operations.
Insertion and replacement use external genomes (subtrees) from benign PDFs.
The original operations generate a lot of different
PDF malware, but not as many different feature inputs to the neural network,
because they don't affect valid Hidost paths in the feature space.
Therefore, we use a trie to index the external benign genomes with valid feature paths.
Given an insertion/replacement point in the PDF malware, we select a genome that shares the same prefix from the trie, to effectively change the input features to the neural network.

Second, we implement a more diverse selection strategy for every generation of the PDF variants.
Diversity is crucial to avoid the evolutionary algorithm being stuck at premature convergence.
We keep all the variants that survived from the previous generation, as in the
original implementation. However, for those variants that are no longer malicious according to the cuckoo sandbox, we find replacement for them
equally by three shares. The first share selects the best historical variant. The second share selects the historical variants with distinct highest
fitness scores, since distinct scores show that the variants explored different mutation paths in the search space. The last share
randomly selects from a pool of historical variants from the last four generations as well as the original seed,
since randomness helps the search process explore more diverse paths that could lead to the solution.


\subsubsection{Experiment Setup}

Following Xu et al.~\cite{xu2016automatically}, we use the same parameters for the attack:
48 variants as the size of the population for each generation, a maximum of 20 generations
per PDF for each round, 0.0 fitness stop threshold, and 0.1 mutation rate. We select
four PDFs with the most benign classification scores as the external
benign genomes, containing a total of 42,629 PDF objects.

\subsection{Trace Length of Evolutionary Attacks}
\label{section:Trace Length of Evolutionary Attacks}

\begin{figure}[t]
    \centering
    \scalebox{0.3}{\includegraphics{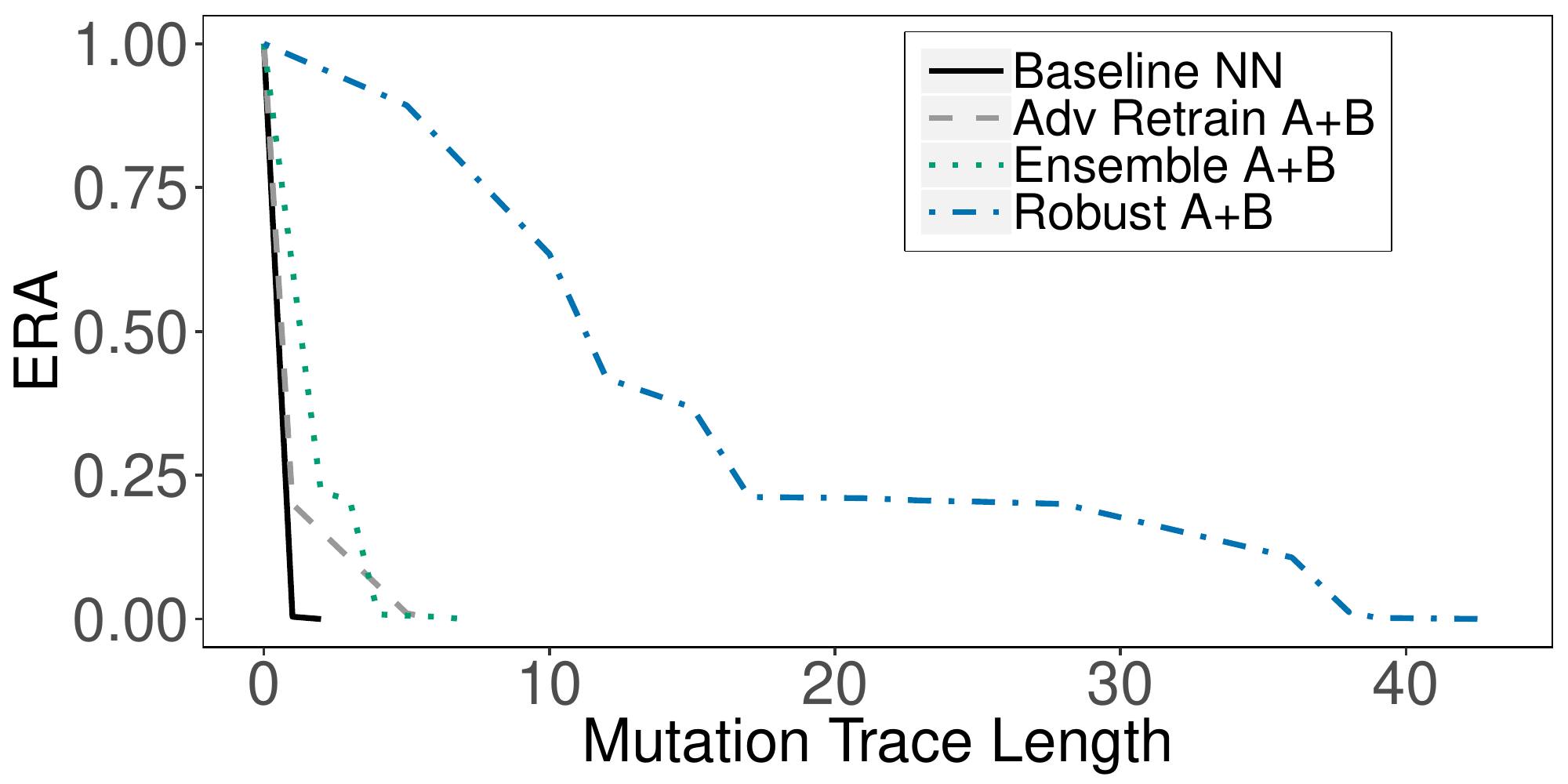}}
    \caption{Trace length for enhanced evolutionary attack.
    }
   \label{fig:era_trace_nonadaptive}
\end{figure}

\paragraph{Enhanced Evolutionary Attack.}
We measure the shortest mutation trace lengths
needed to generate all the PDF variants. Figure~\ref{fig:era_trace_nonadaptive} shows how ERA decreases as the length of mutation trace increases. The Baseline NN
is the easiest to evade. One mutation drops the ERA to 0.4\%. Two mutations are
sufficient to evade the Baseline NN for all PDF seeds. The Adv Retrain A+B and Ensemble A+B models
perform better than the Baseline NN. They can be evaded by up to 6 and 7 mutations respectively.
The Robust A+B requires most number of mutations to evade compared to all other models.
Robust A+B model has higher VRA in property C than the Ensemble A+B model (Table~\ref{tab:vra}), which
further increases the mutation trace length to evade the model.
The attack needs 15 mutations to succeed in 63\% of PDF seeds, and 43 mutations to
succeed in all seeds for Robust A+B.
Since we can verify that Robust A+B is robust against basic building block attack operations (insertions and deletions), unrestricted attacks consisting of the building block operations are harder to succeed.

\begin{figure}[t]
    \centering
    \scalebox{0.3}{\includegraphics{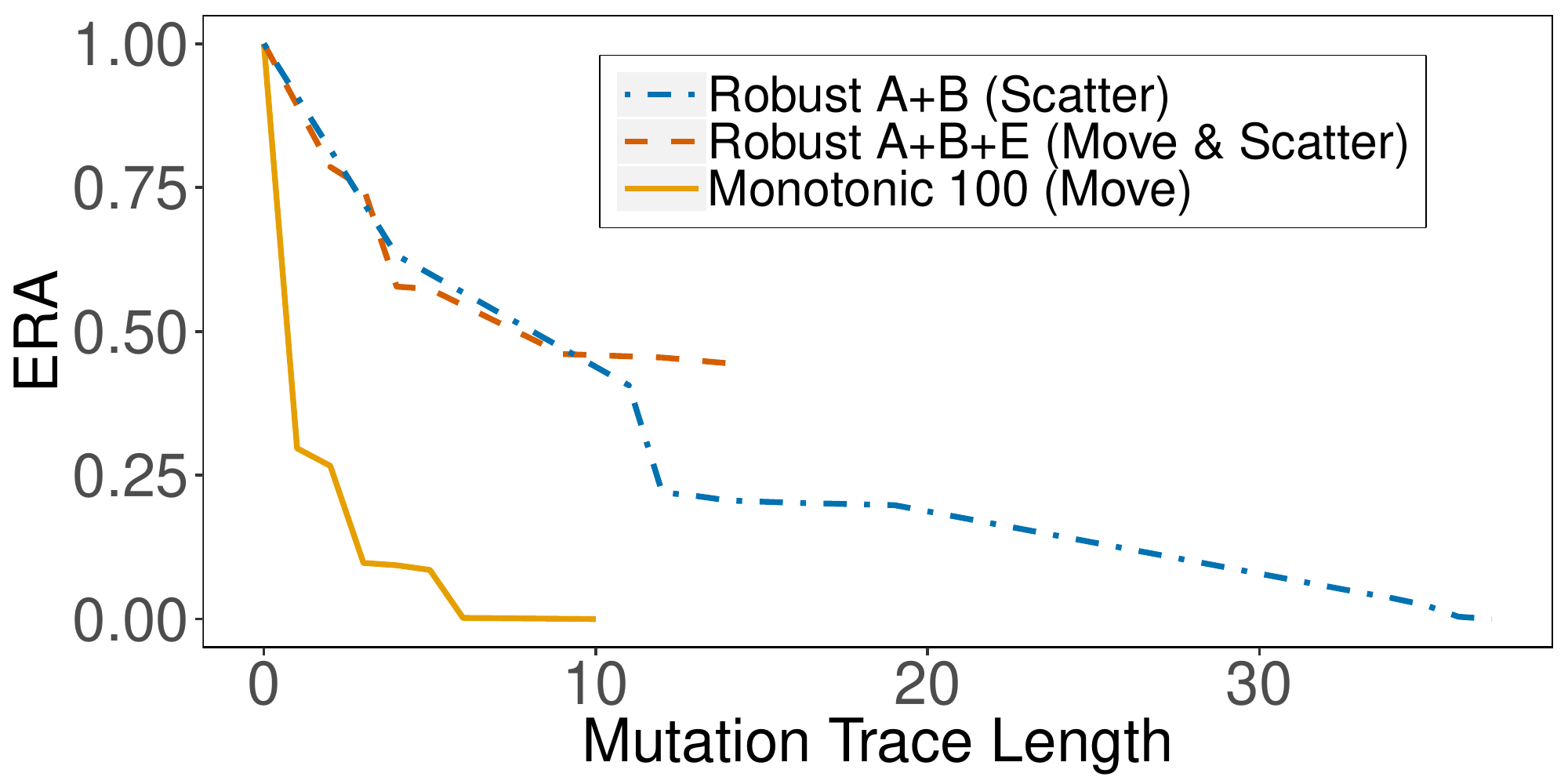}}
    \caption{Trace length for adaptive evolutionary attacks.
    }
   \label{fig:era_trace_adaptive}
\end{figure}

\paragraph{Adaptive Evolutionary Attacks.}
We measure the mutation trace length to evade the three models: Monotonic 100, Robust A+B,
and Robust A+B+E. The move exploit attack is very
effective against the Monotonic 100 model. A single move operation can
decrease the ERA of the model to 29.70\%, e.g., moving
the exploit from \texttt{/Root/OpenAction/JS} to \texttt{/Root/Pages/Kids/AA} .
On the other hand, the scatter attack
can reduce the median mutation trace length needed to evade Robust A+B
from 11 to 8 compared to the nonadaptive version (Figure~\ref{fig:era_trace_nonadaptive}).
The move and scatter combination attack can reduce the ERA of Robust A+B+E to 44\%
with up to 14 mutations.

\subsection{Move Exploit}
\label{section:Move Exploit}

We use the static analyzer peepdf~\cite{peepdf} and manual analysis to identify the following exploit triggers.

\begin{itemize}
\itemsep0em
\item \texttt{/Root/Pages/Kids/AA}
\item \texttt{/Root/Names/JavaScript/Names}
\item \texttt{/Root/OpenAction/JS}
\item \texttt{/Root/StructTreeRoot/JS}
\end{itemize}

The move mutation operation identifies whether the PDF variant has object under one of the paths,
then randomly selects one of the target paths to move the object to. Compared to the reverse mimicry,
the move operation works much better by preserving many references (e.g., function names)
in the same PDF.


\end{document}